\documentclass[12pt]{article}
\pdfoutput=1
\usepackage{cite}
\usepackage{showlabels}
\usepackage{enumerate}
\usepackage{ifpdf}
\usepackage{ae}
\usepackage[T1]{fontenc}
\usepackage[ansinew]{inputenc}
\usepackage{amsmath}
\usepackage{relsize}
\usepackage{amssymb}
\usepackage{comment}
\usepackage{graphicx}
\usepackage{color}
\definecolor{darkblue}{cmyk}{0.9,0.9,0,0}
\definecolor{darkgreen}{rgb}{0,0.55,0}
\usepackage[colorlinks=true,linkcolor=darkblue,citecolor=darkblue,urlcolor=darkblue]{hyperref}
\usepackage{epsfig}
\usepackage{epstopdf}
\usepackage{mathtools}
\usepackage{graphicx}
\DeclareMathOperator*{\dDisc}{dDisc}

\makeatletter
\long\def\@makecaption#1#2{
  \vskip\abovecaptionskip
  \sbox\@tempboxa{{\captionfonts #1: #2}}
  \ifdim \wd\@tempboxa >\hsize
    {\captionfonts #1: #2\par}
  \else
    \hbox to\hsize{\hfil\box\@tempboxa\hfil}
  \fi
  \vskip\belowcaptionskip}
\makeatother

\def\as{{\a' s\o 4}}
\def\c{\cite}
\def\nonan{\mathcal{A}^{(g=1)}_{\rm non-analytic}}
\def\cA{\mathcal{A}}
\def\cMt{\widetilde{\mathcal{M}}}
\def\cR{\mathcal{R}}
\def\cM{\mathcal{M}}
\def\cH{\mathcal{H}}
\def\zb{\bar z}
\def\cG{\mathcal{G}}
\def\cN{\mathcal{N}}
\def\cA{\mathcal{A}}

\def\c{\cite}
\def\1{{\rm 1-loop}}

\def\vs{\vskip .1 in}

\def\G{\Gamma}
\def\p{\partial}
\def\o{\over}
\def\g{\gamma}
\def\D{\Delta}
\def\rar{\rightarrow}

\def\eqr{\eqref}
\def\dDisc{{\rm dDisc}}
\def\O{{\cal O}}

\def\la{\langle}
\def\ssec{\subsection}
\def\sssec{\subsubsection}
\def\sec{\section}
\def\i{\infty}
\def\foot{\footnote}
\newcommand{\es}[2] {\begin{equation} \label{#1} \begin{split} #2 \end{split} \end{equation}}
\newcommand{\e}[2] {\begin{equation} \label{#1} #2 \end{equation}}

\newcommand{\beq}{\begin{equation}}
\newcommand{\eeq}{\end{equation}}
\newcommand{\beqy} {\begin{eqnarray}}
\newcommand{\eeqy} {\end{eqnarray}}
\newcommand{\bsmat}{\begin{smallmatrix}}
\newcommand{\esmat}{\end{smallmatrix}}
\newcommand{\bmat}{\begin{matrix}}
\newcommand{\emat}{\end{matrix}}

\def\({\left(}
\def\){\right)}
\def\[{\left[}
\def\]{\right]}

\def\<{\langle}
\def\>{\rangle}

\def\a{\alpha}
\def\b{\beta}
\def\g{\gamma}
\def\G{\Gamma}

\def\D{\Delta}
\def\z{\zeta}

\def\l{\lambda}

\def\s{\sigma}
\def\vs{\vskip .1 in}

\usepackage{varioref}
\usepackage{makeidx}
\makeindex

\usepackage[english]{babel}
\usepackage{caption}
\usepackage{subfig}

\usepackage{tabularx}
 
\begin{document}

\thispagestyle{empty}

\renewcommand{\thefootnote}{\fnsymbol{footnote}}
\setcounter{page}{1}
\setcounter{footnote}{0}
\setcounter{figure}{0}
\begin{titlepage}

\begin{center}

\vskip 2.3 cm 

\vskip 5mm

{\LARGE \bf Genus-One String Amplitudes from \vs Conformal Field Theory
}
\vskip 0.5cm

\vskip 15mm
\centerline{ Luis F. Alday$^{1/c}$, Agnese Bissi$^{1/\lambda}$ and Eric Perlmutter$^{\alpha'}$}
\bigskip
\centerline{\it $^{1/c}$ Mathematical Institute, University of Oxford,} 
\centerline{\it Woodstock Road, Oxford, OX2 6GG, UK}
\vs
\centerline{\it $^{1/\lambda}$ Department of Physics and Astronomy, Uppsala University,}
\centerline{\it Box 516, SE-751 20 Uppsala, Sweden}
\vs
\centerline{\it $^{\alpha'}$ Walter Burke Institute for Theoretical Physics,}
\centerline{\it  Caltech, Pasadena, CA 91125, USA}

\end{center}

\vskip 2 cm

\begin{abstract}
\noindent We explore and exploit the relation between non-planar correlators in ${\cal N}=4$ super-Yang-Mills, and higher-genus closed string amplitudes in type IIB string theory. By conformal field theory techniques we construct the genus-one, four-point string amplitude in AdS$_5\times S^5$ in the low-energy expansion, dual to an ${\cal N}=4$ super-Yang-Mills correlator in the 't Hooft limit at order $1/c^2$ in a strong coupling expansion. In the flat space limit, this maps onto the genus-one, four-point scattering amplitude for type II closed strings in ten dimensions. Using this approach we reproduce several results obtained via string perturbation theory. We also demonstrate a novel mechanism to fix subleading terms in the flat space limit of AdS amplitudes by using string/M-theory. 

\end{abstract}

\end{titlepage}

\tableofcontents
\numberwithin{equation}{section}

\setcounter{page}{1}
\renewcommand{\thefootnote}{\arabic{footnote}}
\setcounter{footnote}{0}

 \def\nref#1{{(\ref{#1})}}

\newpage

\section{Introduction and summary} 

This paper uses analytic methods of the conformal bootstrap to construct non-planar correlators in $\cN=4$ super-Yang-Mills (SYM), and to relate them to detailed features of perturbative type II closed string amplitudes. Our computations focus on two interrelated aspects.  

The first is the direct construction of the one-loop/genus-one, four-point string amplitude in AdS$_5\times S^5$ in the low-energy expansion. This is holographically dual to the $\cN=4$ SYM four-point function of the lowest half-BPS operator, in the `t Hooft limit at $\O(1/c^2)$ and in a $1/\l$ expansion. In the flat space limit, this maps onto the genus-one, four-point scattering amplitude for type II closed strings in $\mathbb{R}^{10}$, which we call $\cA^{(g=1)}$. We will develop the systematic expansion of this and related half-BPS four-point functions, and give explicit low-orders results. In the flat space limit, we will match the $\cN=4$ SYM correlator to terms in the low-energy expansion of $\cA^{(g=1)}$ constructed from supergravity, $\cR^4$ and $\p^4\cR^4$ vertices. We also perform a match of some results to all orders in $\a'$, including the forward limit of the discontinuity of $\cA^{(g=1)}$, which we derive on the string theory side using existing technology.

The second is a new insight into the interpretation of subleading terms in the flat space limit of AdS amplitudes, and how to fix them using string/M-theory. 

\sssec*{Background}

Four-particle amplitudes in type IIB string theory admit a double expansion: a genus expansion in different topologies in powers of $g_s$, and a low energy expansion in powers of $\alpha'$. For strings on AdS$_5 \times S^5$ one can study this problem by considering holographic correlators in ${\cal N}=4$ SYM, in a double expansion around large central charge $c$ and large `t Hooft coupling $\lambda$. These are correlators of protected chiral primary operators ${\cal O}_p$, dual to Kaluza-Klein (KK) scalars on $S^5$, of dimension $\Delta=p$ and $SU(4)_R$ irrep $[0,p,0]$. The simplest such operator is  ${\cal O}_2$, the superconformal primary of the stress tensor multiplet. We will focus on the four-point function $\la \O_2\O_2\O_2\O_2\rangle$ at $\O(1/c^2)$ in the $1/\l$ expansion, and the matching of its flat space limit to the genus-one, four-point closed string amplitude in the $\a'$-expansion. 

At the planar level, stringy corrections appear as local quartic vertices in the tree-level AdS effective action. The origin of the stringy corrections to the $\cN=4$ SYM correlator is the $S^5$ dimensional reduction of the low-energy expansion of the type IIB action. For instance, quartic terms of schematic form $\p^{2k}\cR^4$, where $\cR$ is the 10d Riemann tensor, generate quartic vertices in AdS$_5$ for all KK components of $\cR$. These translate to polynomial amplitudes in Mellin space, and to linear combinations of so-called $D$-functions in position-space \cite{DHoker:1999kzh,Heemskerk:2009pn,Penedones:2010ue,Alday:2014tsa}. Thus, in the context of AdS$_5$ string theory, $\a'$ corrections appear as polynomial corrections to meromorphic tree-level Mellin amplitudes and, via the holographic relation
\e{}{\a' =L_{\rm AdS}^2/\sqrt{\l}~,}
to the $1/\l$ expansion of the $\cN=4$ SYM Mellin amplitude. In the crossing context, these polynomial corrections are sometimes referred to as ``truncated'' solutions.

At $\O(1/c^2)$ in CFT (one-loop in AdS), amplitudes may be determined by a kind of ``AdS unitarity method.'' This idea -- introduced in \cite{Aharony:2016dwx}, and further developed in \cite{Alday:2017xua} -- computes the one-loop amplitude essentially as a square of the tree-level amplitude. This is made manifest in large spin perturbation theory \cite{Alday:2016njk} and the elegant Lorentzian inversion formula \cite{Caron-Huot:2017vep}, in which CFT correlators are determined, modulo certain low-spin data, by their double-discontinuity (``dDisc''). In particular, dDisc of the one-loop correlator is determined completely by tree-level data. This was leveraged in \cite{Alday:2017vkk}  to compute the full CFT data for the one-loop correlator $\la \O_2\O_2\O_2\O_2\rangle$ at infinite $\l$.\footnote{A proposal for the full correlator was given in \cite{Aprile:2017bgs}. The same CFT data should follow from that proposal.}

\sssec*{Stringy corrections to non-planar correlators}

Our goal here is to incorporate stringy corrections to the one-loop amplitude in AdS$_5 \times S^5$ and, via the flat space limit, to recover genus-one string amplitudes in 10d. We will determine the dDisc of $\la \O_2\O_2\O_2\O_2\rangle$ at $\O(1/c^2)$ to several orders in the $1/\l$ expansion and, from this, extract the physical content of the amplitude using Lorentzian inversion and the flat space limit. From the bulk perspective, we are computing the dDiscs of the one-loop, four-point scattering amplitude for type IIB closed strings in AdS$_5 \times S^5$ in the low-energy expansion. Because all stringy corrections involve quartic vertices, the $1/\l$ expansion of the one-loop correlator is dual to a sum of the box function (one-loop supergravity) plus a tower of four-point triangle and bubble diagrams in AdS$_5$.\foot{This statement is precise modulo non-1PI diagrams, as explained in Section \ref{genus-one}.} These are degenerations of the non-perturbative (in $\a')$ one-loop closed string amplitude:
\vs
\begin{figure}[h]
\includegraphics[scale=0.54]{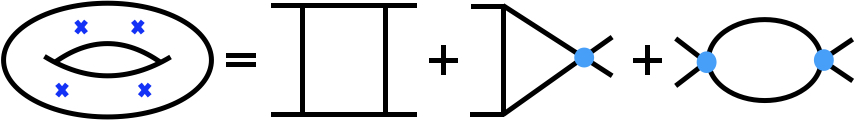}
\centering
\label{fig:torus}
\end{figure}
\vs
\noindent This picture may be thought of as living in $\mathbb{R}^{10}$ or AdS$_5\times S^5$. 

Let us summarize the computation. The correlator $\la \O_2\O_2\O_2\O_2\rangle$ is fixed by a single function of cross-ratios which we call $\cH(z,\zb)$. We will be computing the dDisc of its genus-one term, $\cH^{(g=1)}(z,\zb)$. dDisc$(\cH^{(g=1)})$ is completely determined by the term proportional to $\log^2z$, which is in turn fixed by the square of the tree-level anomalous dimensions, $\g^{(g=0)}$, of $SU(4)_R$ singlet double-trace operators $[\O_2\O_2]_{n,\ell}$, of schematic form
\e{}{[\O_2\O_2]_{n,\ell} = \O_2\p^{2n}\p_{\mu_1}\ldots \p_{\mu_\ell}\O_2 - (\text{traces})}
$\g^{(g=0)}$ is a function of $\l$, admitting an expansion
\e{gamexp}{\g^{(g=0)} \approx \g^{(g=0|\rm sugra)} + \sum_{k=0}^\i\l^{-(3+k)/2}\g^{(g=0|\p^{2k}\cR^4)}}
The superscript refers to the 10d $\p^{2k}\cR^4$, which generates, via dimensional reduction, order $\l^{-(3+k)/2}$ corrections to the AdS$_5$ effective action. In the $1/\l$ expansion, the precise expression for dDisc$(\cH^{(g=1)})$ is a sum of powers of $1/\l$ times sums of the form 
\e{tdef}{T^{\rm x|y}(z,\bar z) \equiv \frac{1}{8} \sum_{n,\ell} a^{(0)}_{n,\ell}\la \g^2\rangle^{\rm x|y}_{n,\ell}\,g_{n,\ell}(z,\bar z)}
The $a^{(0)}_{n,\ell}$ are squared OPE coefficients of mean field theory (MFT), $g_{n,\ell}(z,\zb)$ are the superconformal blocks corresponding to exchange of $[\O_2\O_2]_{n,\ell}$, and
\es{gamnotn}{\la \g^2\rangle^{\rm x|y}_{n,\ell}&\equiv\langle \gamma^{(g=0|\rm x)}\gamma^{(g=0|\rm y)} \rangle_{n,\ell} ~,\quad \text{where ${\rm x,y} = {\rm sugra}$ or $\p^{2k}\cR^4$~.}}
Each term in the expansion may be viewed as computing the dDisc of AdS triangle or bubble diagrams with the appropriate quartic vertices:
\es{}{T^{\,{\rm sugra}|\p^{2k}\cR^4}(z,\zb) \quad &\Leftrightarrow \quad \text{dDisc(AdS$_5$ triangles)}\\
T^{\p^{2k}\cR^4|\p^{2k'}\cR^4}(z,\zb) \quad &\Leftrightarrow \quad \text{dDisc(AdS$_5$ bubbles)}}

The ``amplitudes'' $T^{\rm x|y}$ in \eqr{tdef} will be our main focus. We will compute them explicitly for various cases involving sugra, $\cR^4$ and $\p^4\cR^4$ vertices. Based on this, we make an ansatz for the transcendentality structure of $T^{\rm x|y}$ in \eqr{claim}. The ansatz is simple, involving weight-one functions only, and quite restrictive: indeed, upon specifying the order of the vertices, a basis of solutions can be found. (See Section \ref{special}.) This prescription \eqr{claim} is one of our main results. 

As a technical remark, the computation requires incorporating $1/\l$ corrections into a mixing problem among families of unprotected double-trace operators $[\O_p\O_p]_{n,\ell}$. At $c=\i$, these operators have $\D_{n,\ell} = 2p+2n+\ell$, so the operators $[\O_2\O_2]_{n,\ell}$ are degenerate with $[\O_p\O_p]_{n-(p-2),\ell}$. This diagonalization is the meaning of the brackets in \eqr{tdef}. The mixing problem has been solved recently at $\l=\i$ and to leading order in $1/c$ \cite{Alday:2017xua,Aprile:2017bgs,Aprile:2017xsp,Aprile:2018efk}. As shown in these works, this requires knowledge of the correlators $\langle {\cal O}_2 {\cal O}_2 {\cal O}_p {\cal O}_p \rangle$ to $\O(1/c)$. That this mixing problem arises at one-loop can be seen heuristically via cutting AdS box diagrams involving $\phi_p$ on the internal lines:
\begin{figure}[h]
\includegraphics[scale=0.32]{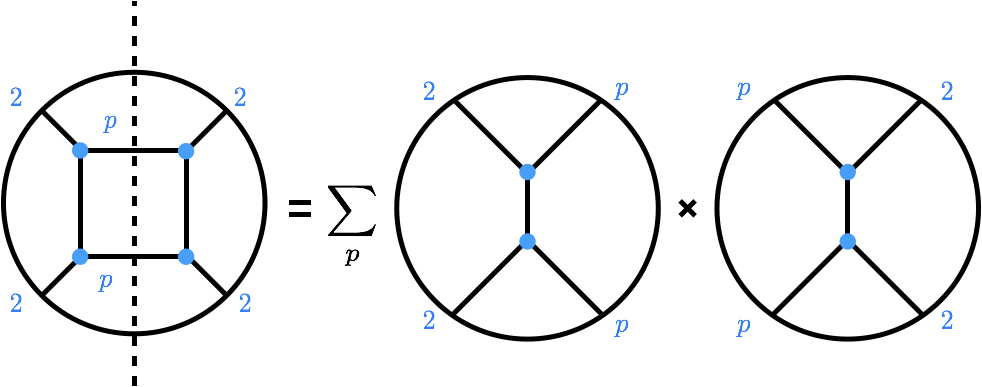}
\centering
\label{fig:box22pp}
\end{figure}
\vs
\noindent where the tree-level diagrams represent the correlators $\langle {\cal O}_2 {\cal O}_2 {\cal O}_p {\cal O}_p \rangle$ in the supergravity approximation. In the present work we extend these results to include $1/\l$ corrections. Again solving the mixing problem at $\O(1/c)$, we must now include the ``truncated'' solutions to the crossing equations which correspond to the quartic vertices in AdS$_5$. Such contributions to mixing can then be depicted as follows:
\vs
\begin{figure}[h]
\includegraphics[scale=0.33]{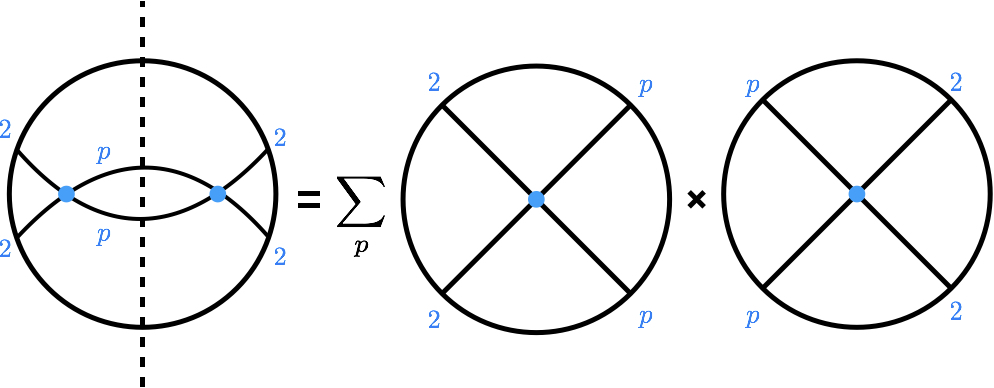}
\centering
\label{fig:bubble22pp}
\end{figure}
\noindent  and similarly for the triangle diagrams. 

Having computed these double-discontinuities, we then turn to extract interesting physical information.  

\sssec*{Anomalous dimensions}
The most natural piece of data are $1/\lambda$ corrections to the anomalous dimensions to $\O(1/c^2)$. This can be directly obtained from the  dDisc of each contribution by the inversion formula, or equivalently large spin perturbation theory. A remarkable feature of these results is the presence of simple poles at certain values of the spin. This implies that the one-loop anomalous dimensions induced by stringy corrections diverge linearly for these values of the spin. This is a CFT manifestation of the UV divergences of one-loop AdS diagrams \cite{Aharony:2016dwx}. In AdS, these divergences are cured by local counterterms, of exactly the same form as the quartic interactions that appear at tree-level. The dimension of these counterterms dictates the maximum spin they can cure, and is related to the degree of the divergence. For each triangle and bubble diagram we show that the values of the spin for which we have poles are exactly the ones expected from the above perspective.  

\sssec*{Flat space limit}

In any CFT with a string/M-theory dual, the leading terms of a Mellin amplitude in the limit $s,t\rar\i$ may be determined by equating the result with the appropriate 10d or 11d flat space string/M-theory scattering amplitude \c{Penedones:2010ue} (see also \c{Susskind:1998vk,Polchinski:1999ry,Okuda:2010ym}). For $\cN=4$ SYM, this relates the non-planar correlator to the genus-one, type IIB closed string amplitude in $\mathbb{R}^{10}$.\foot{It is known that type IIA and IIB four-point scattering amplitudes in $\mathbb{R}^{10}$ are equal through genus four  \cite{Berkovits:2006vc}.} This amplitude, $\cA^{(g=1)}$, is given by an integral of a specific modular function over the fundamental domain of $SL(2,\mathbb{Z})$ \c{Green:1981yb}. The $\a'$ expansion was studied in a series of works \c{Green:1997as,Russo:1997mk,Green:1998by,Green:1999pu,Green:1999pv,Green:2005ba,Green:2006gt}, most systematically in \c{Green:2008uj}. At low orders in the $\a'$ expansion, transcendentality of the coefficients permits an unambiguous split\foot{This is the conclusion of \c{Green:2008uj}, see e.g. Section 4.3. However, adopting a form of transcendental grading in which logarithms of Mandelstam invariants have unit weight, the analytic and non-analytic pieces of the amplitude possess terms of equal weight at low orders. This will not affect our matching between AdS and flat space amplitudes. We thank Eric D'Hoker for raising this issue.} into analytic and non-analytic pieces,
\e{}{\cA^{(g=1)} \propto \cA^{(g=1)}_{\rm analytic}(s,t) + \cA^{(g=1)}_{\text{non-analytic}}(s,t)}
where $s,t$ are 10d Mandelstam invariants.

The analytic piece can be thought of as regulating the one-loop UV divergences of 10d supergravity augmented by the higher-derivative quartic vertices of string theory. We will show how the $\cN=4$ SYM one-loop correlator -- in particular, the pattern of UV divergences exhibited by the anomalous dimensions described earlier -- reflects the precise functional form of $ \cA^{(g=1)}_{\rm analytic}$.

More interesting is the non-analytic piece. In the flat space limit, the {\it double-discontinuity} of the one-loop AdS amplitude becomes the {\it discontinuity} of the 10d amplitude \cite{Alday:2017vkk}:
\e{ddiscflat}{\dDisc(\cH^{(g=1)}) \xrightarrow[\text{flat space limit}] {} \text{Disc}(\cA^{(g=1)})~.}
By taking this limit, we generate predictions for the discontinuities of $\cA^{(g=1)}$ involving sugra, $\cR^4$ and $\p^4\cR^4$ vertices. Using previous results of \cite{Green:2008uj}, we find a match. 

We also use our CFT methods to compute the functional form of certain discontinuities to all orders: first, any term in the $\a'$ expansion of $\cA^{(g=1)}$ involving at least one $\cR^4$ vertex; and second, the complete discontinuity of $\cA^{(g=1)}$ in the limit of forward scattering ($t\rar 0$). We independently derive these results, and determine the actual coefficients, using the string theory techniques of \cite{Green:2008uj}. (See \eqr{discr4string} and \eqr{fwddisc}.) To our knowledge, these expressions have not appeared elsewhere. 
    
    \sssec*{Flat space limit: subleading order}
    
In taking the flat space limit, we run into an interesting open question for holography. Subleading terms at large $s,t$ represent ``finite size corrections'' due to AdS curvature, and are not accessible using naive application of the flat space limit. One would like to know whether these subleading terms -- indeed, the full AdS amplitudes themselves -- may be recast as certain scattering observables in the higher-dimensional string/M-theory and, if so, which ones. 
 
One of our main observations is that subleading terms in the $s,t\rar\i$ limit of {\it tree-level} AdS Mellin amplitudes may actually be fixed by constructing the one-loop AdS amplitude, and matching its flat space limit to a {\it one-loop} string/M-theory amplitude. The basic point is that since the one-loop amplitude is essentially the square of the tree-level amplitude by AdS unitarity, the subleading terms in the tree-level amplitude feed into the one-loop amplitude. Then by matching the latter to the string/M-theory one-loop amplitude, these subleading terms can be at least partially fixed. For the case of $\la \O_2\O_2\O_p\O_p\rangle$ at $\O(1/c)$, the first such subleading term appears at $\O(\l^{-5/2})$, where the Mellin amplitude takes the form\foot{At $\O(\l^{-3/2})$, the leading stringy correction, the amplitude is just a constant which can be matched using the flat space limit and the known 10d $\cR^4$ vertex, see \cite{Goncalves:2014ffa} and Appendix \ref{appb}.}
\e{}{\b (s^2+t^2+u^2) + \b_1+\b_2 s}
where $(\b, \b_1, \b_2)$ are functions of $p$. Only $\b$, given in \eqr{alphabeta}, can be fixed by matching onto the Virasoro-Shapiro amplitude at $\O(\a'^5)$. By matching to the genus-one string amplitude at $\O(\a'^5)$ we fix $\b_2=2p(p-2)(p+1)_4$, and reduce $\b_1$ to two constants.

\vs

\sssec*{Organization}

In Section \ref{generalities} we set up the problem, introduce the tree-level amplitudes in the $1/\l$ expansion, and explain where the subleading terms at large $s,t$ come from in terms of the AdS$_5\times S^5$ reduction. In Section \ref{genus-one} we construct the dDisc of the one-loop amplitudes to the first several orders in $1/\l$, reveal their transcendental structure, and use this to parameterize the coefficients of subleading terms in the tree-level correlator at $\O(\l^{-5/2})$. In Section \ref{data} we make contact with the type II genus-one string amplitudes. We relate their analytic parts to the structure of one-loop anomalous dimensions and UV divergences. Taking the flat space limit of our dDiscs, we reproduce the discontinuities of various terms in the genus-one string amplitudes and constrain the subleading coefficients $\b_1(p)$ and $\b_2(p)$ given above. We end with a handful of open problems, while various Appendices supplement the main text. 
    
\section{Generalities and Tree-Level Solutions}
\label{generalities}
 
Our object of study is the four-point function of $\O_2$, the superconformal primary in the stress tensor multiplet of $\cN=4$ SYM. $\O_2$ is a rank-two symmetric traceless tensor of $SO(6)_R \approx SU(4)_R$. Contracting its R-symmetry indices with polarization vectors $y_i$ obeying the null condition $y_i\cdot y_i=0$, we introduce the index-free four-point function
\begin{equation}
\langle {\cal O}_2(x_1,y_1){\cal O}_2(x_2,y_2){\cal O}_2(x_3,y_3){\cal O}_2(x_4,y_4) \rangle = {(y_1\cdot y_2)^2(y_3\cdot y_4)^2\o x_{12}^4 x_{34}^4}\sum_{\cal R} Y^{\cal R}(\sigma,\tau) {\cal G}^{\cal R}(z,\bar z)
\end{equation}  
where the sum runs over $SU(4)_R$ representations ${\cal R} \in [0,2,0] \times [0,2,0]$. We have introduced cross-ratios in position space
\begin{equation}
z \bar z = \frac{x_{12}^2 x_{34}^2}{x_{13}^2x_{24}^2},~~~(1-z)(1-\bar z) = \frac{x_{14}^2 x_{23}^2}{x_{13}^2x_{24}^2},~~~~~~~x_{ij}=x_i-x_j
\end{equation}
and polarization space,
\e{polcross}{ \sigma\equiv\frac{(y_1\cdot y_3)(y_2\cdot y_4)}{(y_1\cdot y_2)(y_3\cdot y_4)}\,,\qquad \tau\equiv\frac{(y_1\cdot y_4)(y_2\cdot y_3)}{(y_1\cdot y_2)(y_3\cdot y_4)} \,.
 }
The $Y^{\mathcal{R}}(\s,\tau)$ are $SO(6)$ harmonics which may be found in \cite{Nirschl:2004pa}. We will work in the Lorentzian regime, where $z,\bar z$ are independent complex variables. (For the physical correlator, they are real variables.) Superconformal Ward identities  \cite{Nirschl:2004pa,Dolan:2004mu} allow to write all ${\cal G}^{\cal R}(z,\bar z)$ in terms of a single function ${\cal G}(z,\bar z) \equiv \cG^{\bf 105}(z,\zb)/(z\zb)^2$, where the irrep ${\bf 105} \equiv [0,4,0]$. Under the crossing transformation $z \leftrightarrow 1-\bar z$, this satisfies the relation
\begin{equation}
((1-z)(1-\bar z))^2 {\cal G}(z,\bar z) - (z \bar z)^2 {\cal G}(1-\bar z,1-z) + ((z \bar z)^2-((1-z)(1-\bar z))^2)+\frac{z \bar z-(1-z)(1-\bar z)}{c}=0
\end{equation}  
where the central charge $c=(N^2-1)/4$. See\cite{Beem:2016wfs} for a detailed discussion. The contribution to ${\cal G}(z,\bar z)$ from protected intermediate operators, belonging to short multiplets, can be computed exactly and is denoted by ${\cal G}_{\rm short}(z,\bar z)$. We then split
\begin{equation}
{\cal G}(z,\bar z)={\cal G}_{\rm short}(z,\bar z)+{\cal H}(z,\bar z)
\end{equation}  
where ${\cal H}(z,\bar z)$ carries the dynamically non-trivial information and admits a decomposition in superconformal blocks,
\begin{equation}\label{blockexp}
{\cal H}(z ,\bar z) = \sum_{\Delta,\ell} a_{\Delta,\ell} g_{\Delta,\ell}(z,\bar z)~,
\end{equation}  
with squared three-point coefficients $a_{\D,\ell}$. The sum runs over superconformal primaries in long multiplets, of dimension $\Delta$ and (traceless symmetric) Lorentz spin $\ell$. The superconformal blocks are given by
\begin{equation}
g_{\Delta,\ell}(z,\bar z)= (z \bar z)^{\frac{\Delta-\ell}{2}}\, \frac{z^{\ell+1} F_{\frac{\Delta+\ell+4}{2}}(z)F_{\frac{\Delta-\ell+2}{2}}(\bar z) -\bar z^{\ell+1} F_{\frac{\Delta+\ell+4}{2}}(\bar z)F_{\frac{\Delta-\ell+2}{2}}( z) }{z-\bar z}
\end{equation}
where 
\e{}{F_\beta(z) \equiv {}_2F_1(\beta,\beta,2\beta;z)}
is the standard hypergeometric function. ${\cal G}_{\rm short}(u,v)$ is independent of the coupling constant $\lambda$ and is $1/c$ exact \cite{Beem:2016wfs}. In this paper, we will study $\cG(z,\zb)$ at $\O(1/c^2)$; so for our purposes $\cH(z,\zb)$ obeys the homogeneous crossing equation
\e{crossingH}{\cH(z,\zb) = \left({z\zb\o (1-z)(1-\zb)}\right)^2\cH(1-\zb,1-z)\qquad (\O(1/c^2))~.}

In the 't Hooft limit, CFT observables admit an expansion in powers of $1/c$ times functions of the 't Hooft coupling $\l$. In perturbation theory around strong coupling, this becomes a double expansion in $1/c$ and $1/\l$. ${\cal H}(z,\bar z)$ admits a double expansion of the form\foot{At $\O(1/c^2)$ and beyond, there are also $\log\l$ terms. Their existence is implied by the presence of logarithmic threshold terms in the genus-one string amplitude. We will determine the discontinuities of these logs from our CFT results. See Section \ref{sec42} for further discussion.}
\begin{eqnarray}
\label{doubleexpansion}
{\cal H}(z,\bar z) &= {\cal H}^{(0)}(z,\bar z) + c^{-1} \left( {\cal H}^{(g=0)}_{\rm sugra}(z,\bar z) + {\lambda^{-3/2}}{\cal H}^{(g=0)}_{1}(z,\bar z) + {\lambda^{-5/2}}{\cal H}^{(g=0)}_{2}(z,\bar z)  +\cdots \right) \nonumber\\
& + c^{-2}\left( {\cal H}^{(g=1)}_{\rm sugra}(z,\bar z) + {\lambda^{-3/2}}{\cal H}^{(g=1)}_{1}(z,\bar z) + {\lambda^{-5/2}}{\cal H}^{(g=1)}_{2}(z,\bar z)  +\cdots \right)+\cdots
\end{eqnarray}  
${\cal H}^{(0)}(z,\zb)$ is the MFT contribution, while $ {\cal H}^{(g=0)}_{\rm sugra}(z,\bar z)$ is the well known supergravity result \c{Arutyunov:2000py,Dolan:2001tt},\foot{As proven in \c{Alday:2017vkk}, both ${\cal H}^{(0)}(z,\zb)$ and ${\cal H}^{(g=0)}_{\rm sugra}(z,\bar z)$ follow from the structure of singularities as $\bar z \to 1$ in the crossing equation.}
\begin{eqnarray}
 {\cal H}^{(g=0)}_{\rm sugra}(z,\bar z)= - (z \bar z)^2 \bar D_{2,4,2,2}(z,\bar z)
\end{eqnarray}  
The precise powers of $\l$ appearing are inferred from the type IIB string amplitudes. We will give further detail about these in Section \ref{data}.

In strong coupling perturbation theory, the only single-trace operators with finite conformal dimensions are the half-BPS operators $\O_p$. The long operators contributing to $\cH(z,\zb)$ to $\O(1/c^2)$ in the superconformal block decomposition \eqr{blockexp} are the double-trace operators $[\O_p\O_p]_{n,\ell}$. Their scaling dimensions admit an expansion analogous to \eqr{doubleexpansion},
\begin{equation}
\Delta_{n,\ell} = 4+2n+\ell + \frac{1}{c} \left( \gamma^{(g=0|{\rm sugra})}_{n,\ell}+\frac{1}{\lambda^{3/2}} \gamma^{(g=0|\cR^4)}_{n,\ell} + \cdots \right) + \cdots
\end{equation}
and likewise for the squared three-point coefficients $a_{n,\ell} \equiv C^2_{22[pp]_{n,\ell}}$. 
For later convenience we quote the leading-order result for the anomalous dimension
\e{}{ \gamma^{(g=0|{\rm sugra})}_{n,\ell} = - \frac{\kappa_n}{(\ell+1)(\ell+6+2n)}~, \quad \text{where } \kappa_n=(n+1)_4~.}
with $(n+1)_4=\Gamma(n+5)/\Gamma(n+1)$ being the ascending Pochhammer symbol.
 
 As will be clear in the next section, in computing the solutions to $\O(1/c^2)$ we will be forced to consider more general correlators $\langle {\cal O}_2 {\cal O}_2 {\cal O}_p {\cal O}_p \rangle$. The structure of these correlators is almost identical to $\langle {\cal O}_2 {\cal O}_2 {\cal O}_2 {\cal O}_2 \rangle$: in the direct channel $22 \rar pp$, the correlator can again be decomposed into six $SU(4)_R$ representations, and again the superconformal Ward identities determine all six channels in terms of a single function. The dynamically non-trivial information arises from double-trace unprotected operators and is encoded in ${\cal H}_p(z,\zb)$, which admits a double expansion analogous to (\ref{doubleexpansion}).  For general $p$,
\begin{eqnarray}
 {\cal H}^{(g=0)}_{p,{\rm sugra}}(z,\bar z)= -\frac{p}{2\Gamma(p-1)}(z\bar z)^p \bar D_{p,p+2,2,2}(z,\bar z)
\end{eqnarray}  
Note that for $p\neq 2$ crossing relates ${\cal H}_{p}(z,\bar z)$ to a different correlator, so that the relation (\ref{crossingH}) is not satisfied.

\ssec{Mellin space}

We will sometimes use the Mellin space approach to these amplitudes and their stringy corrections. The Mellin representation of the above correlators $\cH_p(z,\zb)$ is defined as\foot{$\cM$ is the ``reduced'' amplitude in the parlance of \cite{Rastelli:2016nze,Rastelli:2017udc}, who call it $\cMt$. Likewise, $u_{\rm here} = \widetilde u_{\rm there}$. How to recover the rational parts of position space amplitudes is discussed in \cite{Rastelli:2017udc}.}
\es{hmellin}{{\cal H}_p(z,\bar z) = & \int_{-i \infty}^{i \infty} \frac{ds dt}{(4\pi i)^2} (z \bar z)^{s/2} ((1-z)(1-\bar z))^{t/2-(p+2)/2} \,\Gamma_{pp22}\,{\cal M}_p(s,t) }
where
\e{}{\Gamma_{pp22} \equiv \Gamma\left(\frac{2p-s}{2} \right) \Gamma\left(\frac{4-s}{2} \right) \Gamma\left(\frac{p+2-t}{2} \right)^2\Gamma\left(\frac{p+2-u}{2} \right)^2}
with $s+t+u=2p$. The crossing conditions simply read
\begin{equation}\label{mcross}
{\cal M}_p(s,t) = {\cal M}_p(s,u),~~~~ {\cal M}_2(s,t) = {\cal M}_2(t,s)
\end{equation}
The supergravity solutions take a very simple form
\begin{equation}\label{mpsugra}
 {\cal M}_{p,{\rm sugra}}(s,t)=\frac{4 p}{\Gamma(p-1)} \frac{1}{(s-2)(t-p)(u-p)}
\end{equation}
which indeed can be seen to satisfy the crossing conditions. In Appendix \ref{appb}, we explain how to take the flat space limit of these Mellin amplitudes and the subsequent relation to type IIB S-matrix elements. 

\subsection{Structure of genus zero solutions} \label{sub.2.2}
Let us now discuss stringy corrections to the supergravity result \eqr{mpsugra}. The case $p=2$ was addressed in \cite{Alday:2014tsa} but the generalisation to arbitrary $p$ is straightforward, following the above rules and imposing the crossing condition \eqr{mcross}. For $p=2$, the solutions are spanned by the basis of monomials 
\e{219}{\sigma_2^m\sigma_3^n,\quad \text{where } \sigma_p \equiv s^p+t^p+u^p~.}
%
$\sigma_2^m\sigma_3^n$ gives rise to double-trace data for spins $\ell\leq L= 2(m+n)$. 
 Generalizing to $p\neq 2$, $\lbrace\sigma_2^m\sigma_3^n\rbrace$ no longer forms a basis, as we can construct more general solutions which obey $t\leftrightarrow u$, but not $s\leftrightarrow t$, crossing symmetry. 
  
While crossing symmetry alone cannot fix the overall coefficient of a solution, conformal Regge theory \c{Costa:2012cb} and unitarity imply that polynomial amplitudes are suppressed by powers of the higher spin gap scale $\D_{\rm gap}$, with the number of powers determined by dimensional analysis \c{Heemskerk:2009pn,Camanho:2014apa,Caron-Huot:2017vep,Meltzer:2017rtf,Afkhami-Jeddi:2016ntf,Kulaxizi:2017ixa,Costa:2017twz}. In the context of string theory in AdS, $\sqrt{\a'} \sim 1/\D_{\rm gap}$. This implies that a $(2m+3n)$-derivative term in the $\cN=4$ SYM Mellin amplitude, such as $\sigma_2^m\sigma_3^n$, appears multiplied by $\lambda^{-(3/2+m+3n/2 )}$, to leading order in $1/\l$. We can thus parameterize the tree-level amplitudes $\cM_p^{(g=0)}(s,t)$ in the $1/\l$ expansion as
\e{mpg0}{\cM_p^{(g=0)}(s,t) = {p\o \Gamma(p-1)}\left({4\o (s-2)(t-p)(u-p)} + \sum_{m,n=0}^\i \l^{-(3/2+m+3n/2)} \cM_{p|m,n}^{(g=0)}(s,t)\right)}
where 
\e{subpowers}{\cM_{p|m,n}^{(g=0)}(s,t) \propto \s_2^m\s_3^n + \text{subleading powers}}

\noindent The presence of subleading powers will be explained momentarily. It is also useful to organize the expansion in momenta rather than in powers of $1/\l$, whereupon the coefficient of a given term has an infinite expansion in $1/\l$. For $p=2$, for example, where $\s_2^m\s_3^n$ form a basis,
\e{}{\cM_2^{(g=0)}(s,t) = {p\o \Gamma(p-1)}\left({4\o (s-2)(t-p)(u-p)} + \sum_{m,n=0}^\i \s_2^m\s_3^n\l^{-(3/2+m+3n/2)}f_{m,n}(\l)\right)}
where $f_{m,n}(\l)$ has an infinite expansion in non-negative powers of $1/\sqrt{\l}$. 

This structure may be understood from the form of the tree-level AdS$_5$ effective action for KK scalars $\phi_p$ dual to $\O_p$. All polynomial Mellin amplitudes for $\cM_p$ are associated to quartic bulk vertices $\phi_2^2\phi_p^2$. The suppression of 10d derivatives by powers of $\a'$ translates directly into $1/\l$ suppression of quartic vertices in AdS$_5$ after dimensional reduction on $S^5$, where we recall that $L_{S^5} = L_{\rm AdS}$. The leading terms in \eqr{subpowers} come from dimensional reduction of the corresponding 10d vertices $\p^{2k}\cR^4$ (+ superpartners), with $2k=4m+6n$. The subleading terms in \eqr{subpowers} come from higher-derivative terms in 10d which have legs on the $S^5$. Conversely, the leading terms may be fixed by the leading asymptotics in the $s,t\rar\i$ limit of the AdS$_5$ amplitude. 

It is useful to write the first few orders explicitly,
\es{melling0}{{\cal M}^{(g=0)}_p(s,t) = \frac{p}{\Gamma(p-1)} \left( \frac{4}{(s-2)(t-p)(u-p)} + \frac{\alpha}{\lambda^{3/2}} + \frac{1}{\lambda^{5/2}} \left(\beta \sigma_2+ \b_1+\b_2 s \right) +  \O(\l^{-3}) \right)}
%
where $(\a,\b,\b_1,\b_2)$ are constant parameters which may depend on $p$. The term of $\O(\l^{-3/2})$ descends from the 10d ${\cal R}^4$ supervertex, while the terms of $\O(\l^{-5/2})$ descend from the 10d $\partial^4 {\cal R}^4$ supervertex.\foot{In what follows, we will refer to the $\l^{-3/2}$ term as the $\cR^4$ term, etc., even though we are always computing AdS$_5$ amplitudes for scalar fields.} We can fix $\a$ and $\b$ by matching to the Virasoro-Shapiro amplitude in the flat space limit, as done in \cite{Goncalves:2014ffa} for the $\cR^4$ term in $p=2$. This is done in Appendix \ref{appb} using the formula \eqr{treereln}, with the result
\e{alphabeta}{\a = \zeta_3(p+1)_3 ~, \quad \b = {\zeta_5 \o 8} (p+1)_5~,}
where $\zeta_s$ is the Riemann zeta function. However, $\b_1$ and $\b_2$  descend from the 10d $\p^4\cR^4$ supervertex with legs on the $S^5$: they are subleading in the flat space limit, and cannot be fixed by this method alone. One of the aims of this paper is to understand to what extent we can fix such subleading parameters, thus making a precise identification between truncated solutions and quartic vertices in the AdS$_5$ effective action. For future convenience, we redefine 
\e{}{\b_1 \equiv \b_1(p) (p+1)_3~,\quad \b_2 \equiv \b_2(p) (p+1)_4~.}

In space-time, these truncated solutions have a relatively simple structure, involving rational and transcendental functions, of the form
 \begin{equation}
\left. {\cal H}(z,\bar z) \right|_{\s_2^m\s_3^n} = R_0(z,\bar z) + R_1(z,\bar z) \log z \bar z+ R_2(z,\bar z)   \log (1-z)(1-\bar z)+ R_2(z,\bar z)  \Phi(z,\bar z)
\end{equation}
where $\Phi(z,\bar z)$ is the standard one-loop scalar box integral. An important feature of these rational functions is that they have a divergence as $z \to \bar z$, of the form
 \begin{equation}
\s_2^m\s_3^n \to  R_i(z,\bar z) \sim \frac{1}{(z-\bar z)^{13+4m+6n}}
\end{equation}
As discussed in \cite{Gary:2009ae,Heemskerk:2009pn,Okuda:2010ym,Maldacena:2015iua}, this singularity is expected for holographic CFT's with a local bulk dual. It is also directly related to the large $n$ behaviour of the $\gamma_{n,\ell}$ generated by these solutions: $\sigma_2^{m'}\sigma_3^{n'}$ generates $\g_{n,\ell}^{(m',n')}$ with behavior 
\begin{equation}
\g^{(m',n')}_{n,\ell}\sim n^{9+4m'+6n'} \qquad (n\gg1)
\end{equation}
In a general sum of the form
\e{}{f(z,\zb) = \sum_{n,\ell} a^{(0)}_{n,\ell} \psi_{n,\ell} g_{n,\ell}(z,\zb)}
for some $\psi_{n,\ell}$, we expect
 \e{bprule}{\psi_{n\gg 1,\ell} \sim n^\alpha ~~\to~~  f(z,\bar z) \sim \frac{1}{(z-\bar z)^{\alpha+4}} \quad \text{as}~~ z\rar\zb~.}

\section{One-Loop Solutions}
\label{genus-one}
Let us now proceed to construct the tower of one-loop solutions, at $\O(1/c^2)$ in CFT. We will follow closely the strategy of \cite{Alday:2017xua}, where ${\cal H}^{(g=1)}_{\rm sugra}(z,\bar z)$ was constructed. The idea was explained in the introduction: determine the double-discontinuity (dDisc) of the amplitude, and use the Lorentzian inversion formula to extract the full OPE data (and construct the full amplitude if one wishes). 

The dDisc of an amplitude $\cH(z,\zb)$ may be defined as the difference between the Euclidean correlator and its two possible analytic continuations around $\bar z=1$, keeping $z$ held fixed:
\begin{equation}
{\rm dDisc}\,[{\cal H}(z,\bar z)] \equiv {\cal H}(z,\bar z) -\frac{1}{2} {\cal H}^\circlearrowleft(z,\bar z)-\frac{1}{2} {\cal H}^\circlearrowright(z,\bar z). \label{ddisc}
\end{equation}
Note that integer powers of $(1-\bar z)$ times $\log(1-\zb)$ have vanishing dDisc. At strong coupling, all powers of $(1-\zb)$ are indeed integer, because the spectrum consists of $\O_p$ and their composites. Consequently, the full dDisc of the one-loop amplitudes comes from the piece proportional to $\log^2(1-\zb)$. By crossing, which takes $z\rar 1-\zb$, this maps to terms proportional to $\log^2 z$. Hence we are interested in finding this piece of the correlator. The $\log^2z$ terms come exclusively from the squared genus-zero anomalous dimensions. Using the expansion in superconformal blocks,
\begin{equation}
\label{ddg1}
\left. {\cal H}^{(g=1)}(z,\bar z) \right|_{\log^2z}= \frac{1}{8} \sum_{n,\ell} \langle a^{(0)} (\gamma^{(g=0)})^2 \rangle_{n,\ell}\, g_{n,\ell}(z,\bar z) 
\end{equation}
where $g_{n,\ell}(z,\zb)$ stands for the conformal block evaluated at $\Delta=4+2n+\ell$. The anomalous dimension $\gamma^{(g=0)}$ is the full anomalous dimension at $\O(1/c)$, and admits the $1/\l$ expansion in \eqr{gamexp}. We have used the bracket to denote an implicit sum over all operators of approximate twist $4+2n$ and spin $\ell$. This is necessary due to mixing: as noted in the introduction and reviewed in Appendix \ref{appa}, for given quantum numbers $(n,\ell)$, there are $n+1$ nearly-degenerate operators of the same spin $\ell$. 
\begin{equation}
[{\cal O}_2,{\cal O}_2]_{n,\ell},[{\cal O}_3,{\cal O}_3]_{n-1,\ell},\cdots, [{\cal O}_{2+n},{\cal O}_{2+n}]_{0,\ell}.
\end{equation}
The intermediate operators in the conformal block expansion of $\cH(z,\zb)$ are the eigenfunctions $\Sigma^{I}$ of the dilatation operator, where $I=1,\cdots n+1$, and (suppressing all other indices)
\e{}{\langle a \gamma^2 \rangle \equiv \sum_{I=1}^{n+1} a_I^{(0)} \gamma_I^2}
In this section we determine $\dDisc({\cal H}^{(g=1)}(z,\bar z)|_{\log^2z})$ to the first few non-trivial orders in the $1/\l$ expansion by expanding $\gamma_I$ in $1/\l$.

\ssec{Review of one-loop supergravity calculation}

As shown in \cite{Alday:2017xua,Aprile:2017bgs,Aprile:2017xsp}, in order to solve the mixing problem that appears at $\O(1/c^2)$, one needs to consider the family of holographic correlators $\langle {\cal O}_2 {\cal O}_2 {\cal O}_p {\cal O}_p \rangle$ to $\O(1/c)$. In \cite{Alday:2017xua} the leading supergravity result, with no stringy corrections, was considered. The final result for the weighted average $\langle (\gamma^{(g=0|{\rm sugra})})^2 \rangle_{n,\ell}$ is a complicated expression and can be found in \cite{Alday:2017xua}. A remarkable feature is its behaviour for large $n$,
\begin{equation}\label{n11}
\langle (\gamma^{(g=0|{\rm sugra})})^2 \rangle_{n,\ell} \sim n^{11}\qquad (n\gg1)
\end{equation}
Without mixing, the square would instead behave as the square of the supergravity result, namely $\sim n^6$. One can interpret the extra $n^5$ as arising from the presence of the $S^5$ in the gravity dual. Using $\langle (\gamma^{(g=0|{\rm sugra})})^2 \rangle_{n,\ell}$, one can compute the final expression for the above sum, which yields
\es{transcendental}{
\left. {\cal H}^{(g=1)}_{\rm sugra}(z,\bar z) \right|_{\log^2z} &=  R_0(z,\bar z) + R_1(z,\bar z)(\text{Li}_2(z) - \text{Li}_2(\bar z)) + R_2(z,\bar z)(\log^2(1-z) - \log^2(1-\bar z)) \\
&+ R_3(z,\bar z)(\log(1-z) - \log(1-\bar z))+ R_4(z,\bar z)(\log(1-z) + \log(1-\bar z)) }
%
for some rational functions $R_i(z,\bar z)$ which can be found in \cite{Alday:2017vkk}. In terms of AdS, this represents the double-discontinuity of the box diagram. An important feature of these rational functions is that they contain the factor
\begin{equation}
R_i(z,\bar z) \propto \frac{1}{(z-\bar z)^{15}}
\end{equation}
Following the discussion at the end of Section \ref{generalities}, this follows from \eqr{n11} as expected. 

\ssec{Adding stringy corrections}\label{sec32}
We now include higher order terms in $1/\lambda$. Having fixed the truncated solutions for ${\cal H}_p(u,v)$ to $\O(1/c)$, we can compute the averages $\langle (\gamma^{(g=0)})^2 \rangle_{n,\ell}$ in a large $\lambda$ expansion by solving the mixing problem order-by-order, and then plug into (\ref{ddg1}). As the complexity of the computation grows quickly, we focus on the first few orders. This will be enough to understand the systematics of the expansion and will already provide explicit new results. Using the shorthand \eqr{gamnotn}, the $1/\l$ expansion of \eqr{ddg1} is of the form\foot{We will sometimes use superscripts $(m,n)\equiv \s_2^m\s_3^n$, as in \eqr{subpowers}, to distinguish different structures at $\p^{2k}\cR^4$.}
\e{}{\left. {\cal H}^{(g=1)}(z,\bar z) \right|_{\log^2z}= T^{{\rm sugra}|\rm sugra}(z,\bar z) + \sum_{k=0}^\i T^{{\rm sugra}|\p^{2k}{\cal R}^4}(z,\bar z) + \sum_{k=0}^\i \sum_{k'=0}^\i T^{\p^{2k}\cR^4|\p^{2k'}{\cal R}^4}(z,\bar z)}
where $T^{\rm x|y}(z,\bar z) $ was defined in \eqr{tdef}. 

\begin{figure}[t]
\centering
\subfloat{\includegraphics[width = 2.2in]{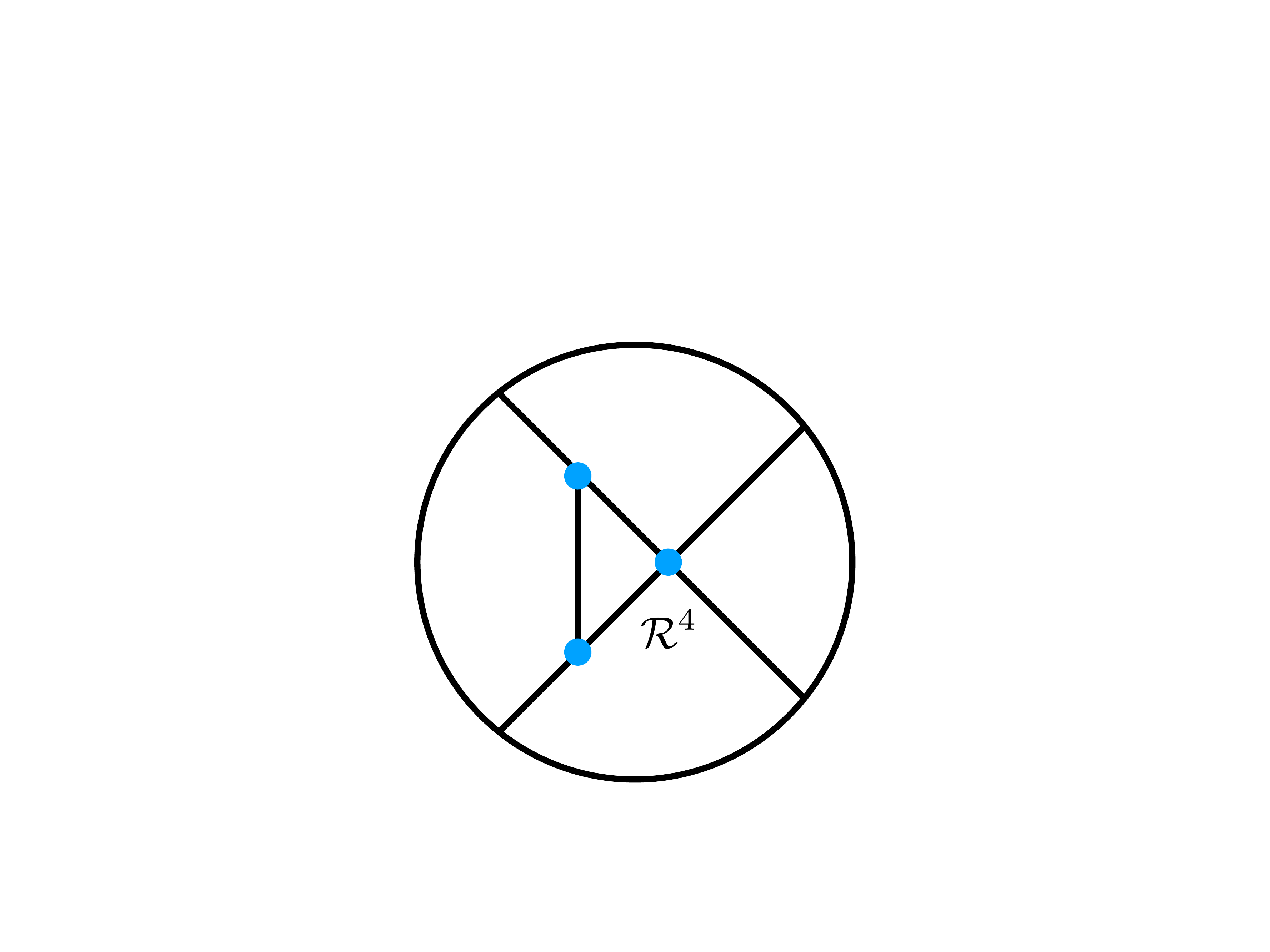}} 
\subfloat{\includegraphics[width = 2.2in]{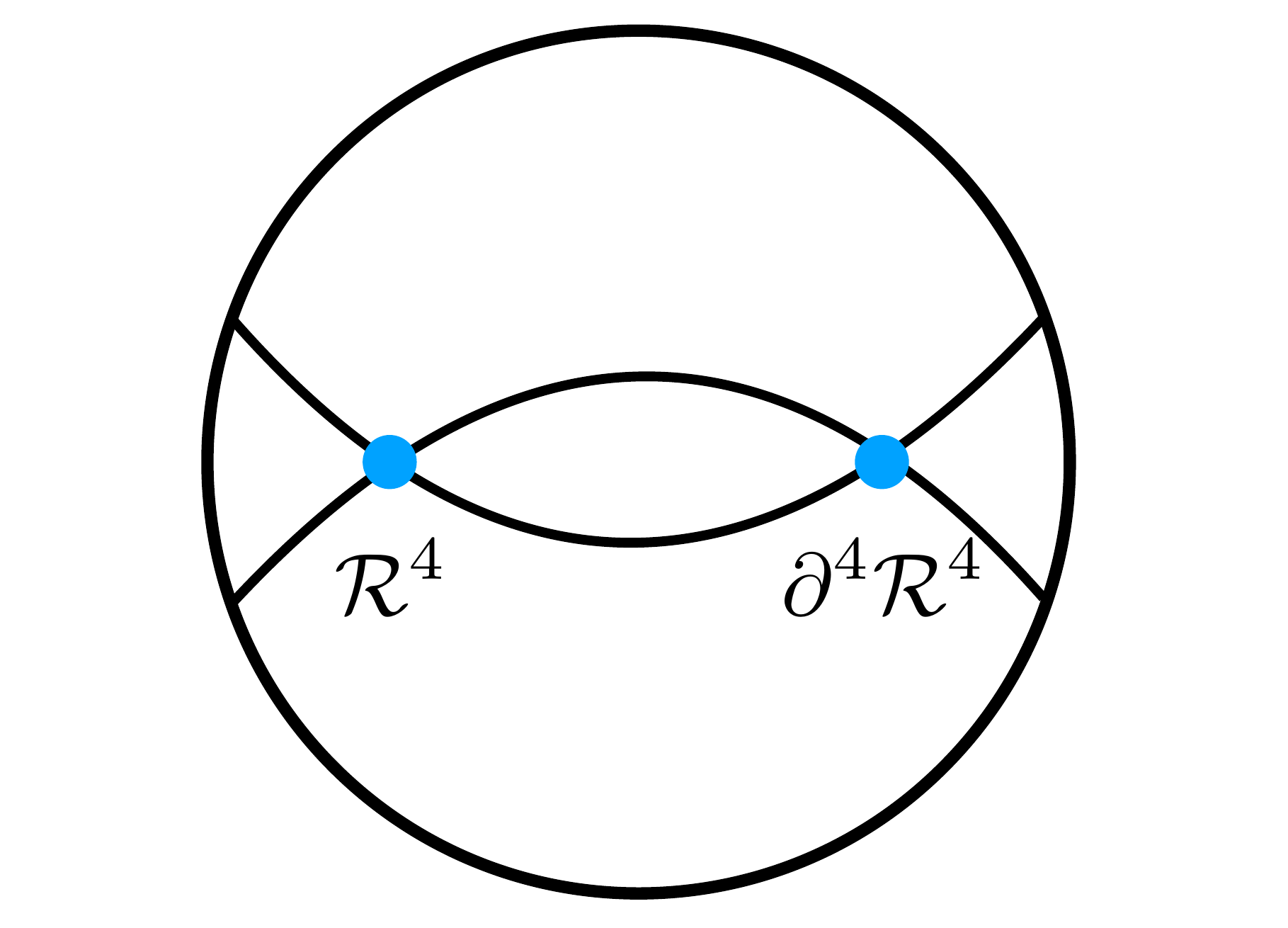}} 
\caption{Two contributions to the genus-one AdS amplitude. The respective sums in \eqr{tsums} compute their dDiscs.}
\label{tribub}
\end{figure}

We will consider the sums involving sugra, $\cR^4$ and $\p^4\cR^4$ vertices:
\es{tsums}{\O(\l^{-3/2}): \quad T^{{\rm sugra}|{\cal R}^4}(z,\bar z)&~\equiv ~\frac{1}{8} \sum_{n,\ell} a^{(0)}_{n,\ell}  \langle \gamma^2\rangle_{n,\ell}^{\rm sugra| \cR^4}g_{n,\ell}(z,\bar z) \\
\O(\l^{-5/2}): \quad T^{{\rm sugra}|\p^4{\cal R}^4}(z,\bar z)&~\equiv ~ \frac{1}{8} \sum_{n,\ell} a^{(0)}_{n,\ell}  \langle \gamma^2\rangle_{n,\ell}^{\rm sugra| \p^4\cR^4}g_{n,\ell}(z,\bar z) \\
\O(\l^{-3}): \quad T^{{\cal R}^4|{\cal R}^4}(z,\bar z)&~\equiv ~ \frac{1}{8} \sum_{n,\ell} a^{(0)}_{n,\ell}  \langle \gamma^2\rangle_{n,\ell}^{\rm \cR^4| \cR^4}g_{n,\ell}(z,\bar z) \\
\O(\l^{-4}): \quad T^{{\cal R}^4|\partial^4 {\cal R}^4}(z,\bar z)&~\equiv ~ \frac{1}{8} \sum_{n,\ell} a^{(0)}_{n,\ell}  \langle \gamma^2\rangle_{n,\ell}^{\rm \cR^4| \p^4\cR^4}g_{n,\ell}(z,\bar z) \\
\O(\l^{-5}): \quad T^{\p^4{\cal R}^4|\partial^4 {\cal R}^4}(z,\bar z)&~\equiv ~  \frac{1}{8} \sum_{n,\ell} a^{(0)}_{n,\ell}  \langle \gamma^2\rangle_{n,\ell}^{\p^4\cR^4| \p^4\cR^4}g_{n,\ell}(z,\bar z) \\}
As explained in the introduction, each term in the expansion may be viewed as computing the dDisc of an AdS triangle or bubble diagram with the appropriate higher-derivative vertices.\foot{There are also vertex corrections and mass and wave function renormalizations. For instance, there exists a bubble vertex correction to tree-level exchange -- in which the bubble has one cubic and one quartic vertex -- which is of the same order in $1/\l$ as a four-point triangle. These non-1PI diagrams are also part of the AdS picture of the stringy corrections being computed here.} We depict the AdS diagrams for two such terms in Figure \ref{tribub}. 

The functions $T$ should have the following properties:
\begin{enumerate}
\item {\it Symmetry under exchange $1 \leftrightarrow 2$.} This is a symmetry of the full correlator and of each superconformal block. It acts on cross-ratios as $(z,\bar z) \to (\frac{z}{z-1},\frac{\bar z}{\bar z-1})$ and maps the piece proportional to $\log^2z$ to itself. This implies
\begin{equation}
T(\frac{z}{z-1},\frac{\bar z}{\bar z-1})  = (1-z)^2(1-\zb)^2 T(z,\bar z)
\end{equation}
\item {\it Absence of terms proportional to $\log^2(1-\bar z)$. }This arises from the fact that the sum over spins is truncated. Hence, it cannot produce a double-discontinuity around $\bar z=1$.
\end{enumerate}
It turns out that these two properties are quite restrictive. If one further assumes that the functions admit a transcendental form analogous to (\ref{transcendental}) this forbids functions of transcendentality higher than one.\foot{This is indeed the case for $T^{{\rm sugra}|{\cal R}^4}$ and $T^{\cR^4|{\cal R}^4}$, as we will see by direct computation below. It will also be borne out in the flat space limit, when we recover the genus-one type II string amplitude in $\mathbb{R}^{10}$.} 

Let us consider the anomalous dimension averages involving the vertex ${\cal R}^4$. From the results in Appendix \ref{appa}, and using $\a=\zeta_3 (p+1)_3$ from \eqr{alphabeta}, we find
\es{r4gammas}{
\langle \gamma^2\rangle_{n,\ell}^{\rm sugra|\cR^4}&=~ \zeta_3\,  \frac{(n+1)^3 (n+2)^4 (n+3)^5 (n+4)^4 (n+5)^3}{720 (2 n+5) (2 n+7)}\,\delta_{\ell,0} \\
\langle \gamma^2\rangle_{n,\ell}^{\cR^4| \cR^4}&=~\zeta_3^2 \,\frac{(n+1)^4 (n+2)^5 (n+3)^7 (n+4)^5 (n+5)^4}{3360 (2 n+5) (2 n+7)}\,\delta_{\ell,0}}
The $p$-dependence  $\a\propto (p+1)_3$ is critical to the rationality of these results. At $n\gg1$,
\begin{eqnarray}
\langle \gamma^2\rangle_{n\gg 1,0}^{\rm sugra|\cR^4} \sim n^{17},~~~~~
\langle \gamma^2\rangle_{n\gg1,0}^{\cR^4| \cR^4}\sim n^{23}
\end{eqnarray}
This agrees with expectations: recalling that the supergravity solution goes like $n^3$, while the first truncated solution goes like $n^9$, so taking into account the extra factor of $n^5$ from mixing indeed yields $17=3+9+5$ and $23=9+9+5$. We can now plug \eqr{r4gammas} into the sums (\ref{ddg1}). The final result has a very simple structure,
\begin{eqnarray}\label{tr4a}
T^{{\rm sugra}|{\cal R}^4}(z,\bar z)&=& \frac{P^{{\rm sugra},{\cal R}^4}_1(z,\bar z)+ P^{{\rm sugra},{\cal R}^4}_0(z,\bar z) \log(1-\bar z) -  P^{{\rm sugra},{\cal R}^4}_0(\bar z,z) \log(1- z) }{(z-\bar z)^{21}} \nonumber\\
T^{\cR^4|{\cal R}^4}(z,\bar z) &=& \frac{P^{{\cal R}^4,{\cal R}^4}_1(z,\bar z)+ P^{{\cal R}^4,{\cal R}^4}_0(z,\bar z) \log(1-\bar z) -  P^{{\cal R}^4,{\cal R}^4}_0(\bar z,z) \log(1- z) }{(z-\bar z)^{27}}
\end{eqnarray}
where $P_i^{{\rm sugra},{\cal R}^4}(z,\bar z),P_i^{{\cal R}^4,{\cal R}^4}(z,\bar z)$ are polynomials of degree 19 and 25, respectively.\footnote{These are available from the authors on request.} The power of $(z-\bar z)$ in the denominator is consistent with the rule \eqr{bprule}. These are complicated polynomials, but we now show how to characterise them and their higher derivative cousins using general considerations. 

\sssec{A basis of special functions}\label{special}
Consider the following sums for generic insertions $\rho_{n,\ell}$
\begin{equation}\label{lsums}
S_L(z,\zb) \equiv \sum_{\ell}^L \sum_{n}a^{(0)}_{n,\ell}  \rho_{n,\ell} g_{n,\ell}(z,\bar z) 
\end{equation}
where $L$ is a non-negative integer. For the problem at hand, the insertion $\rho_{n,\ell}$ corresponds to the averaged squared anomalous dimension; as argued earlier, in the context of stringy corrections we expect this to have the following structure
\begin{equation}\label{lsums2}
S_L(z,\zb) = \frac{R_0(z,\bar z)}{(z-\bar z)^m}+\frac{R_1(z,\bar z) \log(1-\bar z) \pm R_1(\bar z,z) \log(1- z)}{(z-\bar z)^m}
\end{equation}
$m$ and $L$ are non-negative integers, and $R_0(z,\bar z),R_1(z,\bar z)$ are rational functions with simple denominators. The sign in the second term depends on whether $m$ is even or odd, so as to be symmetric under the $z \leftrightarrow \bar z$ symmetry of the superconformal blocks $g_{n,\ell}(z,\zb)$. 

What is the most general form of the insertions $\rho_{n,\ell} $ that leads to this structure, and what are the allowed functions $R_i(z,\zb)$ and the integers $(L,m)$? To be precise, for each $m$ we have searched for solutions where the rational functions $R_i(z,\bar z)$ truncate at some order in a small $z,\bar z$ expansion. This order could in principle be very high, but the correct symmetry under $(z,\bar z) \to (\frac{z}{z-1},\frac{\bar z}{\bar z-1})$ puts an upper bound. By studying the explicit sums over conformal blocks and imposing the above condition, we can count the number of independent solutions, and study their explicit form. The number of solutions depends on $L$.\foot{Although very related, note that this is {\it not} the same problem as the one considered in \cite{Heemskerk:2009pn}. There, the task was to find crossing-symmetric amplitudes formed from conformal blocks and their $\D$-derivatives. The ansatz \eqr{lsums2} is not crossing-symmetric, and is formed out of conformal blocks alone. In our phyiscal problem it represents the dDisc of an amplitude, not a full amplitude. }

At $L=0$, we obtain the following family of solutions, labelled by $q=0,1,\cdots$:
\begin{equation}
\rho^{(0,q)}_{n,0}= \frac{\Gamma (n+q+6)}{(2 n+5) (2 n+7) \Gamma (n-q+1)}
\end{equation}
This obeys $\rho^{(2,q)}_{n\gg1,0} \sim n^{3+2q}$. Denoting the full sums by $S_0^{(q)}(z,\bar z)$, they take the form
\begin{equation}
S_0^{(q)}(z,\bar z) =  \frac{P^{(5+2q)}_1(\bar z,z)}{(z-\bar z)^{7+2q}}+\frac{P^{(5+2q)}_0(z,\bar z)\log(1-\bar z)  - P^{(5+2q)}_0(\bar z,z)\log(1- z)}{(z-\bar z)^{7+2q}} 
\end{equation}
where the $P^{(d)}_i(z,\zb)$ are degree-$d$ polynomials. Explicit results 
are given in Appendix \ref{appc}. The $S_0^{(q)}(z,\zb)$ are related by a differential recursion in $q$.

At $L=2$, we have a new family of solutions, again labelled by $q=0,1,\cdots$:
\begin{eqnarray}
\rho^{(2,q)}_{n,0}&=&\frac{3 (2 q+5) \Gamma (n+q+7)}{(2 n+3) (2 n+5) (2 n+7) (2 n+9) (2 q+9) \Gamma (n-q)} \\
\rho^{(2,q)}_{n,2}&=&\frac{\Gamma (n+q+8)}{(2 n+5) (2 n+7) (2 n+9) (2 n+11) \Gamma (n-q+1)}
\end{eqnarray}
Note that the relative coefficient between the two terms is fixed. This again obeys $\rho^{(2,q)}_{n\gg1,0} \sim \rho^{(2,q)}_{n\gg1,2} \sim n^{3+2q}$. 
Denoting the full sums by $S_2^{(q)}(z,\bar z)$, their general structure is of the form
\begin{equation}
S_2^{(q)}(z,\bar z) = \frac{P^{(9+2q)}_0(z,\bar z)}{\bar z^3 (z-\bar z)^{7+2q}} \log(1-\bar z) -  \frac{P^{(9+2q)}_0(z,\bar z)}{z^3 (z-\bar z)^{7+2q}}  \log(1- z) +   \frac{P^{(7+2q)}_1(\bar z,z)}{z^2 \bar z^2 (z-\bar z)^{7+2q}}
\end{equation}

We conjecture these to be the complete set of solutions at $L=0,2$. The procedure can be carried out at higher $L$ as desired. A generic feature, checked through $L=6$, seems to be that
\e{slzzb}{S_L^{(q)}(z,\zb) \propto (z-\zb)^{-(7+2q)}}
%

\sssec{General prescription}

  With these families of functions at hand, let's now turn to the one-loop stringy corrections. Our claim is that the sums \eqr{tsums} and their higher-derivative partners must be writable as linear combinations of the sums $S_L(z,\zb)$, where $L$ is determined by the derivative order. This follows from the functional ansatz \eqr{lsums2}. This leads to the following general prescription:
\vskip .02 in
{\centerline{\noindent\rule{15cm}{0.8pt}}}

\noindent {\bf Prescription:} For a $\p^{4m+6n}\cR^4$ contribution $(m,n)\equiv \s_2^m\s_3^n$ at one or both vertices, 
\es{claim}{T^{\,{\rm sugra}|(m',n')}(z,\zb) &= \sum_{s=0}^{2(m'+n')}\sum_{q=0}^{7+2m'+3n'} c_{s,q}\, S_s^{(q)}(z,\zb)~,\\
T^{(m,n)|(m',n')}(z,\zb) &= \sum_{s=0}^{s_{\rm max}}\sum_{q=0}^{q_{\rm max}} c_{s,q} \,S_s^{(q)}(z,\zb)}
for some constants $c_{s,q}$, where
\es{sqmax}{s_{\rm max} = 2\times\text{min}(m+n,m'+n')~, \quad q_{\rm max} = 10+2(m+m')+3(n+n')}
The upper bounds on $s$ follow from the discussion below \eqr{219}. The upper bounds on $q$ are determined by power counting (e.g. the growth of $\la \g_{n,\ell}^2\rangle$ at $n\gg1$) and the behavior of solutions at $z=\zb$, under the assumption \eqr{slzzb}.\footnote{Starting from $\partial^{12} {\cal R}^4$ there are multiple structures. At $\partial^{12} {\cal R}^4$, both $\sigma_2^3$ and $\sigma_3^2$ appear. The former has support up to spin 6 and has the schematic form $(\partial_{\mu_1 \mu_2 \mu_3} \mathcal{R}) (\partial_{\mu_4\mu_5 \mu_6} \mathcal{R} ) (\partial_{ \mu_1 \mu_2 \mu_3}\mathcal{R}) (\partial_{\mu_4 \mu_5 \mu_6} \mathcal{R})$, while the latter has support up to spin-4 and has schematic form $(\partial_{\mu_1 \mu_2 \mu_3} \mathcal{R}) (\partial_{\mu_1\mu_4 \mu_5} \mathcal{R}) (\partial_{ \mu_2 \mu_4 \mu_6} \mathcal{R}) (\partial_{\mu_3 \mu_5 \mu_6} \mathcal{R})$, where $\p_{abc}\equiv \p_a\p_b\p_c$.}

For the $\cR^4$ diagrams computed in \eqr{tr4a} one obtains
\begin{eqnarray}\label{tr4}
T^{\,{\rm sugra}|{\cal R}^4}(z,\zb) &= &\z_3\Big(180 S_0^{(0)}(z,\bar z) +3060  S_0^{(1)}(z,\bar z) + \frac{8505}{2}  S_0^{(2)}(z,\bar z) + \frac{2525}{2}  S_0^{(3)}(z,\bar z)\nonumber\\&& + \frac{925}{8}  S_0^{(4)}(z,\bar z) + \frac{153}{40}  S_0^{(5)}(z,\bar z)+ \frac{269}{5760}  S_0^{(6)}(z,\bar z)  + \frac{1}{5760}  S_0^{(7)}(z,\bar z)\Big) \nonumber\\
T^{{\cal R}^4|{\cal R}^4}(z,\zb) &= &\z_3^2\,\Big( \frac{97200}{7}S_0^{(0)}(z,\bar z) + \cdots + \frac{1}{26880}S_0^{(10)}(z,\bar z)\Big) 
\end{eqnarray}
where the explicit $S_0^{(q)}(z,\zb)$ are given in Appendix \ref{appc}. For brevity, we have refrained from writing all terms in $T^{{\cal R}^4|{\cal R}^4}(z,\zb)$, but the structure obeys the ansatz \eqr{claim} with rational coefficients; we have written the $S_0^{(10)}(z,\bar z)$ term explicitly for later use. 

\ssec{$\p^4\cR^4$ and subleading terms in the flat space limit}

Let us now turn our attention to the low-order diagrams involving $\partial^4 {\cal R}^4$. Recall that the tree-level $\p^4\cR^4$ term given in \eqr{melling0} contains two functions, $\b_1(p)$ and $\b_2(p)$, that are not naively determined by the flat space limit of $\cM_p^{(g=0)}$. The solution of the mixing problem for $\la \g^2\rangle$ (presented below using results of Appendix \ref{appa}), and the subsequent sums in \eqr{tsums}, depend on $\beta_1(p)$ and $\beta_2(p)$ in a rather non-trivial way. How do we constrain these functions? 

First, note that the expected large $n$ behaviour of the above contributions enforces $\beta_1(p),\beta_2(p)$ to grow at most as $p^4$ and $p^2$ for large $p$, respectively. 

More powerfully, the claim \eqr{claim} imposes an infinite set of linear and quadratic constraints for $\beta_1(p)$ and $\beta_2(p)$. In order to understand them, let us warmup with the following simpler problem. Consider the truncated solution corresponding to the ${\cal R}^4$ vertex and shift it by a $p$-dependent ambiguity:
\begin{equation}
\alpha \to \alpha + \chi(p)
\end{equation}
and require that contributions including $\chi(p)$ are linear combinations of the family of solutions $S_0^{(q)}(z,\bar z)$. Which constraints does this impose on $\chi(p)$? We obtain a set of linear constraints from the contributions $T^{{\rm sugra}|\chi},T^{\alpha|\chi}$ and a set of quadratic constraints from $T^{\chi|\chi}$. Quite remarkably these constraints imply that $\chi(p)$ has to be an even polynomial in $p$, so that we have the freedom
\begin{equation}
\alpha \to \alpha + \chi_n p^{2n}
\end{equation}
Of course, the correct flat space limit for $\cR^4$ uniquely fixes $\a=\z_3(p+1)_3$.

Let us return to the vertex $\partial^4 {\cal R}^4$. In this case the constraints are much harder to study, but the only solution we were able to find corresponds again to polynomials $\beta_1(p),\beta_2(p)$. We believe this is the most general solution. Recall furthermore that the maximum degree is limited by the large-$n$ behaviour. More precisely, we obtain 
\begin{eqnarray}\label{b1b2}
\beta_1(p) &=& b_1+(40-4b_0) p + b_2 p^2 + (b_0-18)p^3+b_3 p^4\\
\beta_2(p) &=& -\frac{1}{4} (p-2) (b_0 p+2 b_0-8 p)
\end{eqnarray}
where we have also used the condition $\beta_2(2)=0$, which follows from crossing symmetry of the Mellin amplitude for $p=2$. Parameterizing $\b_1(p)$ and $\b_2(p)$ by these polynomials, the diagrams involving $\p^4\cR^4$ take the form
\begin{eqnarray}\label{td4r4}
T^{{\rm sugra}|\partial^4 {\cal R}^4}&=& {\z_5\o 8}\Big(60 (3 b_1+12 b_2+48 b_3+1376)S_0^{(0)}(z,\bar z) +25200S_2^{(0)}(z,\bar z) + \cdots  \nonumber\\
&&+ \frac{-405 b_0+504b_3+10408}{8709120}S_0^{(9)}(z,\bar z) + \frac{1}{241920} S_2^{(9)}(z,\bar z) \Big)\\
T^{ {\cal R}^4|\partial^4 {\cal R}^4}&=& {\z_3\z_5\o 8}\Big(\frac{64800}{7} (3 b_1+12 b_2+48 b_3+1376) S_0^{(0)}(z,\bar z) + \cdots\nonumber\\&& + \frac{-55 b_0+70 b_3+1372}{2661120}S_0^{(12)}(z,\bar z)\Big) \nonumber\\
T^{\partial^4 {\cal R}^4|\partial^4 {\cal R}^4}&=&\left({\z_5\o 8}\right)^2\Big(\frac{10800}{7} (3 b_1+12 b_2+48 b_3+1376)^2 S_0^{(0)}(z,\bar z) +\frac{20321280000}{11} S_2^{(0)}(z,\bar z) + \cdots  \nonumber\\
&&+ \frac{16835 b_0^2-46620 b_0 b_3-747400 b_0+34188 b_3^2+965552 b_3+9167536}{5119994880}S_0^{(14)}(z,\bar z) \nonumber\\
& &+ \frac{1}{997920} S_2^{(14)}(z,\bar z)\Big)\nonumber
\end{eqnarray}
Together with \eqr{tr4}, this determines, up to four constants, all contributions containing the vertices ${\cal R}^4$ and $\partial^4 {\cal R}^4$. 

To summarize, requiring that $\b_1(p)$ and $\b_2(p)$ be consistent with the basis ansatz \eqr{claim} reduces these otherwise-arbitrary functions to four coefficients $\lbrace b_0,b_1,b_2,b_3\rbrace$. Note that $b_0$ and $b_3$, but not $b_1$ and $b_2$, appear in the terms with largest values of $q$.
As we will see in the next section, this will imply that $b_0$ and $b_3$ can actually be fixed using the flat space limit at $\O(1/c^2)$.\foot{It is clear from \eqr{melling0} that for any {\it fixed} $p$, there should be only two undetermined constants at $\O(\l^{-5/2})$. The power of the above analysis is that {\it i)} two constants determine the amplitude for {\it all} $p$, and {\it ii)} $\b_2(p)$ can actually be fixed. This simple $p$-dependence ultimately reflects the symmetries of the $S^5$ which unify the amplitudes for different $p$, as nicely exhibited at the level of tree-level supergravity in the recent work \c{Caron-Huot:2018kta}. It would be interesting to combine the insights of \c{Caron-Huot:2018kta} with the method we are using here at one-loop.} This implies, using \eqr{b1b2}, that $\b_2(p)$ -- which determines the first subleading correction of $\cM_p^{(g=0)}(s,t)$ at large $s,t$ -- may in fact be completely fixed by the flat space limit. The result is given in \eqr{b0b3}--\eqr{betas}.

\section{CFT Data and Genus-One String Amplitudes}
\label{data}
In the introduction it has been mentioned that the double-discontinuity of the correlator contains all the relevant physical information, upon plugging it into the Lorentzian inversion formula \c{Caron-Huot:2017vep}. In this section we exploit this fact.\foot{Due to a number of recent reviews and applications of the Lorentzian inversion formula (e.g. \c{Alday:2017vkk,Simmons-Duffin:2017nub,Liu:2018jhs,Caron-Huot:2018kta}), we refer the reader elsewhere for an exposition, instead confining ourselves to its properties that we will directly use. Our computations are most similar to those of \cite{Alday:2017vkk}.} Noting that 
\e{}{\dDisc((1-\zb)^n\log^2(1-\zb)) = 4\pi^2(1-\zb)^n\quad\text{ for }n\in \mathbb{Z}~,}
the dDisc of our correlator is simply $4\pi^2$ times the coefficient of $\log^2(1-\bar z)$. This coefficient is precisely the definition of our amplitudes $T^{\rm x|y}(z,\zb)$ after applying crossing symmetry to pass to the $t$-channel (where dDisc acts trivially):
\begin{equation}
{\rm dDisc}\,({\cal H}^{(g=1)}(z,\bar z) \big|_{\log^2z}) \supset 4\pi^2 \left(\frac{z \bar z}{(1-z)(1-\bar z)}\right)^2 T^{\rm x|y}(1-\bar z,1-z)
\end{equation}

The crux of this section is the match between our CFT results and the type II closed string amplitude at genus-one. To develop the $\a'$ expansion of the latter, we follow the treatment of \c{Green:2008uj}. The amplitude, $\cA^{10}$, takes the form
\e{a10}{\cA^{10} = \kappa_{10}^2g_s^2{\widehat{K}\o 2^6}\left(\cA^{(g=0)} + 2\pi g_s^2\cA^{(g=1)} + \O(g_s^4)\right)}
$\kappa_{10}^2$ is the gravitational coupling in Einstein frame, $S_{\rm IIB} = (2\kappa_{10}^2)^{-1}\int d^{10}x (R+\ldots)$; $g_s$ is the string coupling; and $\widehat K$ is an overall dimension-eight kinematic factor recalled in Appendix \ref{appb}. The genus-zero amplitude is the Virasoro-Shapiro amplitude \c{Virasoro:1969me},
\e{}{\cA^{(g=0)}(\hat\s_2,\hat\s_3) = {\G(-\a' s/4)\G(-\a' t/4)\G(-\a' u/4)\o \G(1+\a' s/4)\G(1+\a' t/4)\G(1+\a' u/4)}}
which admits an expansion
\es{avs}{\cA^{(g=0)}(\hat\s_2,\hat\s_3) &= {2^6\o \a'^3 stu}\exp\left(\sum_{k=1}^\i {2\z_{2k+1}\o 2k+1}(\a'/4)^{2k+1}(s^{2k+1}+t^{2k+1}+u^{2k+1})\right)\\
&\equiv  \sum_{m=0}^\i\sum_{n=-1}^\i c_{mn}\hat\s_2^m\hat\s_3^n}
Here we follow \cite{Green:2008uj} in using the standard string theory notation $\hat\s_n\equiv (\a'/4)^n( s^n + t^n + u^n)$. The supergravity term is $(m,n)=(0,-1)$, with $c_{0,-1} = 3$. The genus-one amplitude, also known as a function of $\a'$, is a sum of analytic and non-analytic piece,
\e{}{\cA^{(g=1)}(\hat\s_2,\hat\s_3) = \cA^{(g=1)}_{\rm analytic}(\hat\s_2,\hat\s_3) + \cA^{(g=1)}_{\text{non-analytic}}(\hat\s_2,\hat\s_3)}
We will give the explicit form of these pieces in what follows. 

\subsection{Analytic terms: Anomalous dimensions and UV divergences}\label{sec41}
The most important physical observable we can extract from $\dDisc(\cH(z,\zb))$ is the set of anomalous dimensions of the double-trace operators $[\O_2\O_2]_{m,\ell}$.  In Appendix \ref{appd} we present the precise expression extracting the $\O(1/c^2)$ anomalous dimension from $\dDisc(\cH(z,\zb))$, obtained from Lorentzian inversion/large spin perturbation theory. 

As explained in Section \ref{sec32}, the contributions involving vertices ${\cal R}^4$ and $\partial^4 {\cal R}^4$ can be written as linear combinations of the functions $S_0^{(q)}(z,\bar z)$ and $S_2^{(q)}(z,\bar z)$. Hence, a convenient way to organise our computation is by considering each of these functions and finding their contributions to $\g$. This can be readily done using \eqr{gammaddisc}. For leading twist ($n=0$) double-trace operators, we find the following simple answer:
\begin{eqnarray}
S_0^{(q)}(z,\bar z) &\to& \gamma^{(q)}_{0,\ell}= -48 \frac{ \Gamma (q+1)^2 \Gamma (q+3) \Gamma (q+4) \Gamma (-q+\ell+3)}{(\ell+1) (\ell+6) \Gamma (q+\ell+5)} \\
S_2^{(q)}(z,\bar z) &\to& \gamma^{(q)}_{0,\ell}= -288 \frac{ \Gamma (q+1)^2 \Gamma (q+3) \Gamma (q+5) \Gamma (-q+\ell+3)}{(2 q+9) (\ell+1) (\ell+6) \Gamma (q+\ell+5)} 
\end{eqnarray}
An important feature of $\g_{0,\ell}^{(q)}$ given above is the presence of simple poles at $\ell=0,1,\cdots, q-3$. Recalling \eqr{claim} and \eqr{sqmax}, this implies that the one-loop anomalous dimensions induced by stringy corrections diverge linearly for $0\leq \ell \leq q_{\rm max}-3$:
\es{anomdiv}{T^{\,{\rm sugra}|{\rm sugra}}:\quad &\g_{0,\ell}^{(g=1)}\text{diverge for }\ell \leq 1
\\
T^{\,{\rm sugra}|(m',n')}:\quad &\g_{0,\ell}^{(g=1)}\text{diverge for }\ell \leq 4+2m'+3n'
\\
T^{(m,n)|(m',n')}:\quad &\g_{0,\ell}^{(g=1)} \text{diverge for }\ell \leq 7+2(m+m')+3(n+n')}
We have included the pure supergravity loop, computed in \cite{Alday:2017vkk,Aprile:2017bgs}, as a useful benchmark.

The results \eqr{anomdiv} nicely exhibit the CFT picture of AdS UV divergences explained in \cite{Aharony:2016dwx} which we now recapitulate. In AdS, UV divergences are cured by local counterterms whose dimension reflects the degree of divergence. But the counterterm dimension, in turn, determines the maximum spin of the anomalous dimensions it generates ($\ell_{\rm max}=2m+2n$). Therefore, the maximum spin for which anomalous dimensions diverge directly translates into the degree of divergence of the full amplitude. This is manifest above: the spin bound is linear in $q_{\rm max}$, which is determined by the same power counting. One may think of these as UV divergences either in AdS or in the flat space limit. 

More importantly, the results are in accord with the structure of $\cA^{(g=1)}_{\rm analytic}$. Translating \eqr{anomdiv} into the associated bulk counterterms implies that
\es{cterm}{T^{\,{\rm sugra}|{\rm sugra}}:\quad \cA^{(g=1)}_{\rm analytic} &\supset \cR^4\\
T^{\,{\rm sugra}|(m',n')}:\quad \cA^{(g=1)}_{\rm analytic} &\supset c_{m'n'}\p^{6+4m'+6n'}\cR^4\\
T^{(m,n)|(m',n')}:\quad \cA^{(g=1)}_{\rm analytic} &\supset c_{mn}c_{m'n'}\p^{12+4(m+m')+6(n+n')}\cR^4}
where $c_{mn}$ is defined in \eqr{avs}. We now compare this to $\cA^{(g=1)}_{\rm analytic}$. The first few terms of $\cA^{(g=1)}_{\rm analytic}$ are (e.g. (4.43) of \cite{Green:2008uj})  
\e{}{ \cA^{(g=1)}_{\rm analytic}(\hat\s_2,\hat\s_3) = {\pi\o3}\left(1+{\z_3\o 3}\hat \s_3+{97\o 1080} \z_5\, \hat\s_2\hat\s_3 + {1\o 30}\z_3^2\left(\hat\s_2^3+{61\o 36}\hat\s_3^2\right)+\ldots\right)}
From the perspective of the derivative expansion around 10d supergravity, $\cA^{(g=1)}_{\rm analytic}$ regulates UV divergences that arise when computing one-loop amplitudes using the quartic vertices implied by the $\a'$-expansion of the Virasoro-Shapiro amplitude. Thus, both the orders in $\a'$ and the transcendentality of the coefficients in $\cA^{(g=1)}_{\rm analytic}$ can be understood by ``squaring'' the Virasoro-Shapiro amplitude.\foot{For instance, the 1 regulates the quadratic divergence of 10d supergravity; the $\hat \s_3$ regulates the divergence of the one-loop triangle involving a $\cR^4$ vertex; and so on. Likewise, the absence of $\p^4\cR^4$ and $\p^8\cR^4$ terms in $\cA^{(g=1)}_{\rm analytic}$ follows from the absence of $\cR^3$ and $\p^2\cR^4$ terms in $\cA^{(g=0)}$, respectively \cite{Green:2008uj}.} This is precisely the form of \eqr{cterm}.

\subsection{Non-analytic terms: The flat space limit}\label{sec42}

We now take the flat space limit of our amplitudes $T^{\rm x|y}$ and match them to the non-analytic genus-one amplitude, $\nonan$.

\sssec{Flat space limit of dDisc}

In \cite{Alday:2017vkk} a very simple quantitative way to relate the AdS amplitude in the flat-space (i.e. bulk-point) limit to the higher-dimensional amplitude was described: the bulk-point limit of the double-discontinuity of $\cH$ equals the discontinuity of $\cA^{10}$. The picture is summarized by Fig. 7 of that paper. From the CFT perspective, the bulk-point limit is implemented in two steps: encircle $z=0$, then send $z\rar\zb$. Parameterizing this limit as 
 \e{bplim}{z=\bar z+2 x \bar z \sqrt{1-\bar z}\quad \text{with } x \to 0}
 the result is
\begin{equation}\label{disc}
\frac{{\rm dDisc} \left[z\bar z(\bar z-z){\cal H}(z^\circlearrowright,\bar z) \right] }{4\pi^2} \to 2\pi i \times \frac{\Gamma(m)}{(2x)^m} \times g_2(\bar z)
\end{equation}
Then 
\e{}{g_2(\bar z)=-\text{Disc}_s(\cA^{10}(s,t))\quad \text{where } s\rar{1-\zb\o\zb} \text{ and }t\rar1~.}
See  \cite{Alday:2017vkk} for a detailed discussion. 

In this limit, the special functions $S_{L}^{(q)}(z,\bar z)$ have simple behavior:
\es{sdisc}{S_0^{(q)}(z,\bar z)  &\to 2\pi i \,  \frac{\Gamma(6+2q)}{(2x)^{6+2 q}}  \left( -8 \left(\frac{1-\bar z}{\bar z} \right)^{q-3}\right) \\
S_2^{(q)}(z,\bar z)  &\to 2\pi i \,  \frac{\Gamma(6+2q)}{(2x)^{6+2 q}}  \left( -8 \left(\frac{1-\bar z}{\bar z} \right)^{q-5} \frac{6 \left((q+4) \bar z^2+2 (q+5) \bar z+q+4\right)}{(2 q+9) \bar z^2}\right) }
%
Note that the analytic continuation around $z=0$ changes the power of $x$ from the naive guess \eqr{slzzb}. We see that only the functions with $q=q_{\rm max}$ contribute to the flat space limit of the amplitudes $T$. Combining this with \eqr{disc}, \eqr{claim} and the functions $S_L^{(q)}(z,\zb)$, one can read off the functional form of the flat space discontinuity for arbitrary $T^{(m,n)|(m',n')}(z,\zb)$. For later use, we also note that in the limit $\zb\rar 0$, the right-hand side behaves as $\zb^{q-3}$ for both $L=0,2$. This appears to persist for all spins $L$.

For the explicit amplitudes in \eqr{tr4} and \eqr{td4r4}, we find 
\begin{eqnarray}\label{tdiscs}
 g_2^{\rm sugra|\cR^4}(\zb) &=& -{\z_3\o 720}\left(\frac{1-\bar z}{\bar z} \right)^{4}\nonumber\\
 g_2^{\cR^4|\cR^4}(\zb) &=& -{\z_3^2\o 3360}\left(\frac{1-\bar z}{\bar z} \right)^{7}\nonumber\\
 g^{{\rm sugra}|\partial^4 {\cal R}^4}_2(\zb) &=&{\z_5\o 8} \left({45b_0-56 b_3\o 120960}\left({1-\zb\o \zb}\right)^6 - {(1-\zb)^4\o 7560 \zb^6}\left(73 \zb^2-143 \zb+73\right)\right)\nonumber\\
 g^{\cR^4|\partial^4 {\cal R}^4}_2(\zb) &=&-{\z_3\z_5\o 8}\frac{14 (5 b_3+98)-55 b_0}{332640}\left(\frac{1-\bar z}{\bar z} \right)^{9}\\\nonumber
    g_2^{\p^4\cR^4|\partial^4 {\cal R}^4}(\zb) &=&-\left({\z_5\o 8}\right)^2
    \Bigg({455 b_0^2-20 b_0 (63 b_3+1010)+4(231 b_3^2+6524 b_3)\o 17297280}\left({1-\zb\o \zb}\right)^{11}\\&&\qquad\qquad\quad+{(1-\zb)^9\o 4324320\zb^{11}}\left(62044 \zb^2-123672 \zb+62044\right)\Bigg)\nonumber
\end{eqnarray}

We also give two all-orders predictions for the functional form of the discontinuity:
\vs
{\bf i) $\mathbf{(m',n') \times \cR^4}$:} Consider all diagrams involving an $\cR^4$ vertex. Because they can be written in terms of $S_0^{(q)}(z,\zb)$ alone, the answer is simply
\es{discr4}{g_2^{\rm sugra|\cR^4} (\zb) &\propto \left(\frac{1-\bar z}{\bar z} \right)^{4}\\
g_2^{(m',n')|\cR^4} (\zb) &\propto \left(\frac{1-\bar z}{\bar z} \right)^{7+2m'+3n'}}
Similarly, the discontinuities coming from amplitudes involving an $\p^4\cR^4$ or $\p^6\cR^4$ vertex are linear combinations of the $L=0,2$ functions in \eqr{sdisc} with $q=q_{\rm max}$ given in \eqr{sqmax}. 
\vs
{\bf ii) $\mathbf{\zb\rar 0}$:} Consider the limit $\zb\rar 0$ after taking the bulk-point limit \eqr{bplim} (i.e. a ``bulk-point-Regge'' limit). These are the kinematics relevant for comparing to the forward limit of $\cA^{10}$. Assuming that the observation below \eqr{sdisc} is correct, we find 
\es{fwddisccft}{ g_2^{{\rm sugra}|(m'',n'')} (\zb\rar 0) &\propto {\bar z}^{-(7+2m''+3n'')}\\
 g_2^{(m',n')|(m'',n'')}(\zb\rar 0) &\propto{\bar z}^{-(10+2(m'+m'')+3(n'+n''))}}

\sssec{String amplitude}
We now turn to the discontinuity of $\nonan$. In \cite{Green:2008uj}, technology was developed to compute the discontinuity at arbitrary order in $\a'$. Assembling various ingredients there, the formula for the $s$-channel discontinuity is\foot{We have used the relation $2\kappa_{10}^2=(2\pi)^7\a'^4$. The prefactor corrects some typos in \c{Green:2008uj}; in particular, there are factor of two discrepancies among their (4.44), (4.45), (5.27) and Appendix E, that we believe we have fixed below. Our formula is consistent with their (5.27).}
\es{discint}{\text{Disc}_s &\cA^{(g=1)} = -2\pi i \left({\a's\o 4}\right)^7 {1\o 120}\\&\times\sum_{m',n'}\sum_{m'',n''} c_{m'n'}c_{m''n''}\int_0^\pi d\theta \sin^7\theta\int_0^{2\pi} d\phi  \sin^6\phi \,(\hat\s_2')^{m'}(\hat\s_3')^{n'}(\hat\s_2'')^{m''}(\hat\s_3'')^{n''}}
where $\hat\s_i' = \hat\s_i(s,t',u')$ and $\hat\s_i'' = \hat\s_i(s,t'',u'')$ with
\es{stuparams}{t' &= -{s\o 2}(1-\cos\theta)~,\quad u' = -s-t'\\
t'' &= -{s\o2}(1+\cos\theta\cos\rho + \sin\theta\cos\phi\sin\rho)~, \quad u'' = -s-t''\\
\rho &= \arccos \left({t-u\o s}\right)}
The total discontinuity of $\cA^{(g=1)}$ is given by the above plus $t$- and $u$-channel crossings.\foot{These discontinuities arise from logarithmic terms in the amplitude of the form $s^\# \log (s\a'/\mu)$ (plus crossings) for some scale $\mu$ which is determined by a perturbative string theory calculation $\a'$\c{Green:2008uj}. The one-loop CFT amplitude will, using $\a' = L_{\rm AdS}^2/\sqrt{\l}$, have $\O(\log\l)$ terms. The scales $\mu$ will manifest themselves in CFT as truncated solutions to crossing at a given order in $1/\l$, with coefficient $\propto \log\mu$. Conveniently, the flat space limit of our CFT correlator, computed via dDisc, lands on the discontinuity itself.} Though not obviously manifest, the integral is symmetric under $(m',n') \leftrightarrow (m'',n'')$. Note that $\hat \s_2' = \left(\as\right)^2(7+\cos 2\theta)$ and $\hat \s_3' = \left(\as\right)^3 {3\o 4}\sin^2\theta$.

We list some low-lying terms in the $\a'$ expansion, using the same superscript notation as \eqr{tdiscs}:
\es{stringdiscs}{(\text{Disc}_s \cA^{(g=1)})^{{\rm sugra}| \cR^4} &= -2\pi i\times{4\pi \z_3\o 45}\left({\a' s\o 4}\right)^4\\
(\text{Disc}_s \cA^{(g=1)})^{\cR^4| \cR^4} &= -2\pi i\times{2\pi \z_3^2\o 105}\left({\a' s\o 4}\right)^7\\
(\text{Disc}_s \cA^{(g=1)})^{{\rm sugra}| \p^4\cR^4} &= -2\pi i\times{\pi \z_5\o 1260}\left({\a's \o 4}\right)^6\left(87+\left({t-u\o s}\right)^2\right)\\
(\text{Disc}_s \cA^{(g=1)})^{\cR^4| \p^4\cR^4} &= -2\pi i\times{4\pi \z_3\z_5\o 135}\left({\a' s\o 4}\right)^9\\
(\text{Disc}_s \cA^{(g=1)})^{\p^4\cR^4| \p^4\cR^4} &= -2\pi i\times{\pi \z_5^2\o 41580}\left({\a's\o 4}\right)^{11}\left(479+\left({t-u\o s}\right)^2\right)}
where $\text{sugra}| \cR^4 = (0,-1)\times (0,0) + (0,0)\times (0,-1)$, $\cR^4 | \cR^4 = (0,0) \times (0,0)$, and so on.

We can also give the discontinuity to {\it all} orders in two cases. 
\vs
{\bf i) $\mathbf{(m',n') \times \cR^4}$:} The first is for all terms involving a $\cR^4$ vertex. It is easy to see from \eqr{discint} that $(m'',n'')=(0,0)$ has no $\rho$-dependence. Then since $t'\propto s$, the result must be proportional to $s^{7+2m'+3n'}$, with prefactor given by a simple class of trigonometric integrals, which we explicitly evaluate:
\e{discr4string}{(\text{Disc}_s \cA^{(g=1)})^{(m',n')| \cR^4} = -2\pi i \left({\a' s\o 4}\right)^{7+2m'+3n'}\left({3\o 4}\right)^{n'}{\pi \z_3\o 96}c_{m'n'}I_{m',n'}}
where
\es{}{I_{m',n'} &\equiv \int_0^\pi d\theta  \,(7+\cos 2\theta)^{m'} (\sin\theta)^{2n'+7}=\frac{2^{3 m'+1} (2 n'+6)\text{!!}}{(2 n'+7)\text{!!}} \, _2F_1\left(-m',n'+4;n'+\frac{9}{2};\frac{1}{4}\right)}
The integral was evaluated using $\cos2\theta = 1-2\sin^2\theta$ and the binomial expansion.
\vs
{\bf ii) Forward limit:} In the forward limit, $t\rar 0$ for fixed $s$. 
We see from \eqr{stuparams} that in this limit, the parameter $\rho\rar0$, hence $t'' \rar u'=-{s\o 2}(1+\cos\theta)$ and 
\e{}{\hat\s_i''(s,t'',u'') \rar \hat\s_i'(s,t',u')\quad (\text{forward limit})~.}
So the functional form of the discontinuity is identical to the case we just considered involving $\cR^4$ vertices, because the integral boils down to powers of $\s_i'$ only. Assembling factors, the discontinuity  in the forward limit may be written as a sum\foot{In the notation of \c{Green:2008uj}, our result \eqr{fwddisc} gives a closed-form expression for Disc$_s\left(\int_{\mathcal{R}_L}{d^2\tau\o \tau_2^2} f_{an}^{(m,0)}(\tau,\bar\tau)\right)$, specifically its $L$-independent part, where $m_{\rm there} = 7+2(m'+m'')+3(n'+n'')$. We note that in the forward limit, the analytic and non-analytic parts are both controlled by $f_{an}^{(m,0)}(\tau,\bar\tau)$, integrated over different parts of the fundamental domain of $SL(2,\mathbb{Z})$. } 
\begin{eqnarray}\label{fwddisc}
(\text{Disc}_s \cA^{(g=1)})\Big|_{t\rar 0} &=& -2\pi i\,{\pi \o 192} \sum_{m',n'}\sum_{m'',n''}c_{m'n'}c_{m''n''}\left({\a' s\o 4}\right)^{7+2(m'+m'')+3(n'+n'')}\left({3\o 4}\right)^{n'+n''}\nonumber\\&&\times~ I_{m'+m'',n'+n''}
\end{eqnarray}

\sssec{Matching}
Let us now compare the results \eqr{fwddisccft} to \eqr{fwddisc}, \eqr{discr4} to \eqr{discr4string}, and  \eqr{tdiscs} to \eqr{stringdiscs}, recalling that we should set $s\rar(1-\zb)/\zb$ and $t\rar 1$ in the string theory results. To facilitate comparison, we work in units $\a'=4$. 

First we match the functional form of the all-orders results. The $\cR^4$ discontinuity predicted by \eqr{discr4} manifestly matches the discontinuity \eqr{discr4string} computed from perturbative string theory. Next, we match in the forward limit, which corresponds to $\zb\rar 0$. We again see that the powers match between \eqr{fwddisccft} and \eqr{fwddisc}.

We now compare the explicit amplitudes at low orders. Starting with ${\rm sugra}\times \cR^4$, we find a match up to overall normalization. Comparing ratios henceforth, we again find a match for the next term, $\cR^4\times \cR^4$:
\e{}{{g_2^{\cR^4|\cR^4}\o g_2^{\rm sugra|\cR^4}} = {(\text{Disc}_s \cA^{(g=1)})^{\cR^4| \cR^4} \o(\text{Disc}_s \cA^{(g=1)})^{{\rm sugra}| \cR^4}}}
Moving onto $\p^4\cR^4$, we use two of the three amplitudes to fix $b_0$ and $b_3$. Without loss of generality, take these to be ${{\rm sugra}\times \p^4\cR^4}$ and ${\cR^4\times  \p^4\cR^4}$. Demanding equality of the ratios yields
\e{b0b3}{b_0=0~, \quad b_3=-2}
This implies that the genus-zero Mellin amplitude $\cM_p$, given in \eqr{melling0}, has parameters
\es{betas}{\b_1(p) &=b_1+p (p (b_2-2 p (p+9))+40)\\
\b_2(p) &= 2p(p-2)}
Having fixed $b_0$ and $b_3$, we compare $\p^4\cR^4 \times \p^4\cR^4$ and once again find a match,
\e{}{{g_2^{\p^4\cR^4|\p^4\cR^4}\o g_2^{\rm sugra|\cR^4}} = {(\text{Disc}_s \cA^{(g=1)})^{\p^4\cR^4| \p^4\cR^4} \o(\text{Disc}_s \cA^{(g=1)})^{{\rm sugra}| \cR^4}}}
This final match is a strong consistency check of this entire calculation, including the parameterization of $\b_1(p)$ and $\b_2(p)$.

\section{Open problems}

Let us mention some open problems/food for thought:

\begin{itemize}
\item  It would be very interesting to understand the simplest way to fix the subleading terms $\beta_1,\beta_2$ in \eqr{melling0}, which appear in the vertex $\partial^4{\cal R}^4$, by CFT considerations alone.   The determination of $\b_2$ in this paper gives a target for planar integrability studies of four-point functions in $\cN=4$ SYM in the strongly coupled regime. This is an especially non-trivial one because, unlike the $\l^{-3/2}$ correction, it is not fixed by the flat space limit matching to the Virasoro-Shapiro amplitude.

\item Related to the point above, for large but finite $\lambda$ there is an infinite number of intermediate single-trace operators. In \cite{Alday:2016htq} it was shown that the truncated solutions arise naturally from these operators getting heavier and heavier, as $\lambda$ grows. The details of this process depend on the dimensions and OPE coefficients of the single-trace operators. This information should in principle be available from integrability. 

\item More generally in AdS/CFT, it would be fascinating if our procedure could be systematically iterated at successively higher loops, to make a clear map between subleading terms in AdS tree-level amplitudes and higher-dimensional loop-level amplitudes. 

\item We would like to extend these methods to higher genera, where the flat space string amplitude is not known, even in an $\a'$ expansion. In type IIA and IIB, state-of-the-art explicit computations stop at genus two at finite $\a'$ \c{DHoker:2005vch}, and genus three at leading order in $\a'\ll 1$ \c{Gomez:2013sla}. Extending the CFT-inspired computations herein to higher loops would involve several challenging, but conceptually familiar, steps. 

\item It would also be quite interesting to make contact with recent investigations of the underlying modular properties of the genus-one (and higher) string amplitude \c{DHoker:2015gmr,DHoker:2017pvk}, for instance, to find a manifestation of $SL(2,\mathbb{Z})$ modular graph forms in non-planar $\cN=4$ SYM correlators. 

\item We have found the amplitudes $T^{\rm x|y}$ to have a very simple structure, which allowed a match to string theory in the flat space limit. This simple structure arises because we have mixing, which is connected to the R-symmetry of the CFT under consideration. In particular, note that the corresponding anomalous dimensions squared have only single poles, but without mixing they would have double poles. This seems to suggest a much finer constraint on the question of which large $N$ CFT's have a local string theory dual.

\end{itemize}

\section*{Acknowledgements} 
AB is supported in part by the Knut and Alice Wallenberg Foundation under grant KAW 2016.0129. EP is supported by Simons Foundation grant 488657, and by the Walter Burke Institute for Theoretical Physics. LFA and EP thank the Simons Workshop on Analytic Approaches to the Bootstrap for hospitality.

\appendix

\section{Truncated solutions in space-time and results to the mixing problem}\label{appa}
In order to solve the mixing problem encountered in the body of the paper, it is convenient to have the solution ${\cal H}^{g=0}_p(z,\bar z)$ in space time. This can be written in terms of $\bar D$-functions, which admit the following representation

\begin{eqnarray}
\bar D_{\Delta_1 \Delta_2 \Delta_3 \Delta_4}(z,\bar z) = & \int_{-i \infty}^{i \infty} \frac{ds dt}{(4\pi i)^2} |z|^{s} |1-z|^{t} \Gamma(s/2+t/2+\Delta_2)\Gamma(s/2+t/2+\Delta-\Delta_4) \\
& \times  \Gamma(-s/2)\Gamma(-t/2)\Gamma(-t/2+\Delta-\Delta_2-\Delta_3)\Gamma(-s/2-\Delta+\Delta_3+\Delta_4) \nonumber
\end{eqnarray}
where $|z|^2=z \bar z, |1-z|^2=(1-z)(1-\bar z)$ and $\Delta=\frac{1}{2} \sum_i \Delta_i$. The integration contour is chosen such that the poles of each gamma function lie on one side or the other. We will then consider the following solutions in space-time
\begin{equation}
\begin{split} \label{spacetimesol}
{\cal H}^{g=0}_p(z,\bar z) =& \frac{p}{\Gamma(p-1)} |z|^{2p} \left(  \frac{1}{2}\bar D_{p,p+2,2,2}(z,\bar z)+\alpha_0(p) \bar D_{p+2,p+2,4,4}(z,\bar z)+ \alpha_1(p) \bar D_{p+2,p+2,5,5}(z,\bar z)   \right. \\
& \left.+ \alpha_2(p)\left( 1+ |z|^2+|1-z|^2\right) \bar D_{p+3,p+3,5,5}(z,\bar z) + \cdots  \right)
\end{split}
\end{equation}
whose Mellin transform is
\begin{equation}
\begin{split}
{\cal M}_p(s,t) =& \frac{p}{\Gamma(p-1)} \left( \frac{4}{(s-2)(t-p)(u-p)}+ \left(\alpha_0(p)+2 \alpha_1(p)-\frac{\alpha_2(p)}{2}(p^2-4p-4)  \right) \right. \\
& \left. -\frac{\alpha_1(p)}{2} s+ \frac{\alpha_2(p)}{4} \sigma_2+ \cdots \right)~.
\end{split}
\end{equation}
This takes the form (\ref{melling0}) after redefining parameters as
\es{abredefine}{\alpha_0(p) &= \frac{\alpha}{\lambda^{3/2}} +\frac{1}{\lambda^{5/2}} \left( (p^2-4p-4)\beta +\beta_1+4\beta_2 \right)\\
\alpha_1(p) &= -\frac{2 \beta_2}{\lambda^{5/2}}\\
\alpha_2(p) &=  \frac{4\beta}{\lambda^{5/2}}~.}
As already mentioned, the intermediate operators are generically degenerate, meaning that there is more than one double-trace operator with a specific $n$ and $\ell$. Such operators are of the form
\begin{equation}
[{\cal O}_2,{\cal O}_2]_{n,\ell},[{\cal O}_3,{\cal O}_3]_{n-1,\ell},\cdots, [{\cal O}_{2+n},{\cal O}_{2+n}]_{0,\ell},
\end{equation}
and eigenfunctions of the dilatation operator $\Sigma^{I}$ are linear combinations of those. We can choose a normalisation in which these operators are orthonormal\footnote{For the singlet representation, the one relevant here, the degeneracy is lifted completely to order $1/c$, see \c{Aprile:2018efk}, so that mixing can be resolved completely.}, namely $\langle \Sigma^{I} \Sigma^{J}\rangle =\delta^{IJ}$. In order to solve the mixing problem, one needs in principle to consider the family of four point functions $\langle {\cal O}_q{\cal O}_q{\cal O}_p{\cal O}_p\rangle$ at order $c^0$ and $c^{-1}$. From these correlators it is possible to extract the averages $\sum_{I} a^{(0)}_{I}$ and $\sum_{I} a^{(0)}_{I} \gamma_{I}$ from which we build the mixing matrix
\begin{equation}
\begin{split}
 M= \begin{bmatrix}
   \gamma|_{2222} &    \gamma|_{2233} &    \gamma|_{2244} &\dots &   \gamma|_{22pp} \\
     \gamma|_{3322} &  \gamma|_{3333} &  \gamma|_{3344} & \dots &  \gamma|_{33pp}\\
    \vdots\\
     \gamma|_{qq22} &  \gamma|_{qq33} &  \gamma|_{qq44} & \dots &  \gamma|_{qqpp}
  \end{bmatrix}
\end{split}
\end{equation}
where we used the shorthand notation  
\e{}{\gamma|_{qqpp}= \frac{\sum_{I} \left(a^{(0)}_{I,pp}a^{(0)}_{I,qq}\right)^{1/2} \gamma_{I}}{\sum_I\left(a^{(0)}_{I,pp}a^{(0)}_{I,qq}\right)^{1/2}}~. }
To reconstruct $\langle {\cal O}_2{\cal O}_2{\cal O}_2{\cal O}_2\rangle$ at $\O(1/c^2)$, we are interested in $\sum_{I}a^{(0)}_I \gamma_I^2$, which is the element (1,1) of the product $M \cdot M$. To compute this term, instead of fully solving the mixing problem, namely extracting each $a^{(0)}_I$ and $\gamma_I$, we only need one row and one column of the mixing matrix:
\begin{equation}
\begin{split}
 \begin{bmatrix}
  \textcolor{red}{\gamma|_{2222}} &    \textcolor{red}{\gamma|_{2233}} &    \textcolor{red}{\gamma|_{2244}} &\textcolor{red}{\dots} &  \textcolor{red}{\gamma|_{22pp}} \\
     \textcolor{red}{\gamma|_{3322}} &  \gamma|_{3333} &  \gamma|_{3344} & \dots &  \gamma|_{33pp}\\
    \textcolor{red}{\vdots}\\
     \textcolor{red}{\gamma|_{qq22}} &  \gamma|_{qq33} &  \gamma|_{qq44} & \dots &  \gamma|_{qqpp} 
  \end{bmatrix}
\end{split}
\end{equation}
Notice that this simplifies enormously the computations, since we only need to consider correlators of the form $\langle {\cal O}_2{\cal O}_2{\cal O}_p{\cal O}_p\rangle$. Thus we can perform the decomposition in conformal blocks of \eqref{spacetimesol} and we can solve the mixing problem.  At  order $c^0$ we have that averages of squared three point functions $\langle a^{(0)}_{n,\ell}\rangle$, for any $p$, are \cite{Aprile:2017xsp}
\begin{equation}
\frac{24 (\ell+1) n! (\ell+2 n+6) \Gamma^2 (n+3) \Gamma (\ell+n+2) \Gamma^2 (\ell+n+4) \Gamma (n+p+3) \Gamma (\ell+n+p+4)}{p^2 (p+1) \Gamma (n+5) \Gamma (2 n+5) \Gamma (p-1) \Gamma^3 (p) \Gamma (\ell+n+6) \Gamma (2 \ell+2 n+7) \Gamma (n-p+3) \Gamma (\ell+n-p+4)}
\end{equation}

Remarkably, the mixing problem can be solved for arbitrary parameters $\alpha_i(p)$. The final expression has the following form
\begin{equation}\label{mixingsol}
\langle (\gamma^{(g=0)})^2 \rangle_{n,\ell} = \sum_{p=2}^{n+2} \left( f^{\rm sugra}_\ell(n,p)+f^{(0)}_\ell(n,p)\alpha_0(p)+ f^{(1)}_\ell(n,p)\alpha_1(p) + f^{(2)}_\ell(n,p)\alpha_2(p)   +\cdots \right)^2
\end{equation}
where $f^{\rm sugra}_\ell(n,p)\neq 0$ for all values of $\ell$, whereas $f^{(0)}_\ell(n,p)$, $f^{(1)}_\ell(n,p)$ contribute only to $\ell=0$, $f^{(2)}_\ell(n,p)$ to $\ell=0,2$, and so on. For sugra we find, at $\ell=0,2$,
\begin{eqnarray}
f^{\rm sugra}_0(n,p)&=&(n+2)(n+4) p \sqrt{\frac{(n+1) (n+5) \left(p^2-1\right) (n-p+3)}{48 (n+p+3)}}\\
f^{\rm sugra}_2(n,p)&=& p \sqrt{\frac{(n+1)_3 (n+5)_3 \left(p^2-1\right) (n-p+3)_3}{432(n+p+3)_3}}~. \nonumber
\end{eqnarray}
For $f^{(0)}_\ell(n,p)$ and $f^{(1)}_\ell(n,p)$ we find
\begin{eqnarray}
f^{(0)}_0(n,p)&=& \frac{(n+2)^2 (n+3)^3 (n+4)^2 p}{ (2 n+5) (2 n+7)  (p+2)(p+3)} c_1(n,p) \nonumber\\
f^{(1)}_0(n,p)&=& -\frac{n (n+2)^2(n+3)^3 (n+4)^2  (n+6) p}{(2 n+5) (2 n+7) (p+2)(p+3)(p+4)} c_1(n,p) \nonumber
\end{eqnarray}
and zero for all $\ell>0$, while for $f^{(2)}_\ell(n,p)$ we find
\begin{eqnarray}
f^{(2)}_0(n,p)&=& \frac{p(n+2)^2(n+3)^3(n+4)^2 }{2(2 n+3) (2 n+5) (2 n+7) (2 n+9)(p+2)_4} c_1(n,p)h(n,p) \nonumber\\
f^{(2)}_2(n,p) &=& \frac{p(n+2)^{3/2}(n+3)^{5/2}(n+4)^3 (n+5) (n+6)^{3/2}}{3(2 n+5)(2 n+7) (2 n+9)(2 n+11) (p+2)_4} c_1(n,p) \sqrt{(4+n-p)_2 (4+n+p)_2} \nonumber
\end{eqnarray}
and zero for all $\ell>2$. The functions $c_1(n,p)$ and $h(n,p)$ are 
\begin{equation}\nonumber
c_1(n,p)=\sqrt{\frac{(n+1)^3(n+5)^3 (p-1)(n-p+3) (n+p+3)}{48(p+1)}}
\end{equation}
and
\es{}{&~h(n,p)=685 + 756 p + 
  179 p^2 \\&+ (1 + n) (5 + n) (539 + 460 p + 
     86 p^2 + (1 + n) (5 + n) (172 + 13 n (6 + n) + (16 - 9 p) p))~.\nonumber}
Notice that in the solution \eqr{mixingsol}, the index $p$ corresponds to the level of the KK modes that participate in the mixing at twist $n$.

\sec{Flat space limit of AdS$_5 \times S^5$ amplitudes and the type IIB S-matrix}\label{appb}

In this appendix we present the Mellin space version of the 4d $\cN=4$ superconformal Ward identity and the flat space limit formula of the 4d $\cN=4$ Mellin amplitudes. Combining them yields formulas relating the structure of the type IIB four-particle scattering amplitude to the flat space limit of the 4d $\cN=4$ Mellin amplitude. 

In this Appendix only, we use $\cM$ to denote the full Mellin amplitude, and $\cMt$ to denote the reduced amplitude, as in \cite{Rastelli:2017udc}. 

\ssec{Superconformal Ward identity}

Our superconformal Ward identity discussion follows \cite{Rastelli:2017udc}. For now, the following applies to a general four-point correlator $\la \O_{p_1}\O_{p_2}\O_{p_3}\O_{p_4}\rangle$. Define 
\e{}{U=z\zb~, \quad V=(1-z)(1-\zb)}
The Mellin amplitude $\cM(s,t;\s,\tau)$ for the full connected correlator $\cG_{\rm conn}(z,\zb;\s,\tau)$ is related to the reduced Mellin amplitude $\cMt(s,t;\s,\tau)$ for the dynamical function $\cH(s,t;\s,\tau)$ by a difference operator, $\widehat{R}$:
\e{c2}{\cM(s,t;\s,\tau) = \widehat R \circ \cMt(s,t;\s,\tau)}
with
\e{}{\widehat R \equiv \tau + (1-\s-\tau)\widehat V + (\tau^2-\tau-\s\tau)\widehat U + (\s^2-\s-\s\tau)\widehat{UV}+\s \widehat{V^2} + \s\tau\widehat{U^2}}
The hatted powers act as 
\es{}{\widehat{U^mV^n}\circ \cMt(s,t;\s,\tau) &\equiv \cMt(s-2m,t-2n;\s,\tau)\\&\times \left({p_1+p_2-s\o2}\right)_m\left({p_3+p_4-s\o2}\right)_m\left({p_2+p_3-t\o2}\right)_n\\
&\times \left({p_1+p_4-t\o2}\right)_n\left({p_1+p_3-u\o2}\right)_{2-m-n}\left({p_2+p_4-u\o2}\right)_{2-m-n}}
The polarization cross-ratios $(\s,\tau)$ we given in \eqr{polcross}. We are interested in the case $(p_1,p_2,p_3,p_4) = (p,p,2,2)$. Denote the Mellin and reduced Mellin amplitudes by $\cM_p$ and $\cMt_p$, respectively. The inverse Mellin transform of $\cMt_p$, given in \eqr{hmellin}, yields the function $\cH(z,\zb)$ considered in the body of the paper. Note that neither $\cMt_p$ nor $\cH(z,\zb)$ are functions of $\s,\tau$.

\ssec{Flat space limit}
Following the logic of \c{Chester:2018dga} where the AdS$_7 \times S^4$ case was considered, we adapt Penedones' formula \c{Penedones:2010ue} to the case of four-point functions of KK modes with $S^5$ momentum. (See also \c{Chester:2018aca} for the $p=2$ case in AdS$_4\times S^7$.) The result is
\e{}{\lim_{L\rar\i} L (L^5V_5) \,\cM_p(L^2s, L^2t;\s,\tau) = {1\o \Gamma(p)}\int_0^\i d\b\, \b^{p-1}e^{-\b}\cA^{10}_p(2\b s, 2\b t; \s,\tau)}
where $L\equiv L_{\rm AdS} = L_{S^5}$ and $L^5 V_5 = \pi^3$ is the $S^5$ volume. We interpret $\cA^{10}_p$ as the 10d flat space amplitude of four supergravitons, with momenta $k_i$ restricted to a five-plane $\mathbb{R}^5 \simeq$ AdS$_5|_{L\rar\i}$, integrated against the $S^5$ wavefunctions of $\phi_p$ and $\phi_2$ and contracted with $SU(4)_R$ polarization vectors $y_i$. On general grounds \c{Chester:2018dga},
\e{c7}{ \lim_{s,t\rar\i} \cM_p(s,t;\s,\tau) = \cA^{10}_\perp(s,t;\s,\tau)\cdot c(p)}
for some constant function $c(p)$. The amplitude $\cA^{10}_\perp(s,t;\s,\tau)$ is the 10d amplitude in the transverse kinematics $y_i\cdot k_i=0$,
\e{}{(y_1\cdot y_2)^2(y_3\cdot y_4)^2 \cA^{10}_\perp(s,t;\s,\tau) \equiv \cA^{10}|_{k_i\cdot y_i=0}}
These kinematics arise because we are taking the flat space limit of an AdS$_5\times S^5$ amplitude, of modes of the 10d graviton with polarizations along the $S^5$ and momenta in AdS$_5$. 

Recall for what follows that $\cA^{10}$ has the form \eqr{a10}. $\widehat K$ is equivalent to the $t_8t_8\cR^4$ tensor, where $\cR_{\mu\nu\rho\sigma}$ is the linearized Weyl curvature in momentum space, and $t_8$ is the same tensor structure appearing in the $\cR^4$ term in the action. It may be defined as (e.g. \cite{DHoker:2005jhf})
\e{}{\widehat K = ((m_1m_2)(m_3m_4) -  4(m_1m_2m_3m_4)+(\text{perms}))^2\,,}
where
\e{}{m_i^{\mu\nu} \equiv \z_i^{[\mu} p_i^{\nu]}~, \quad (m_im_j) \equiv m_i^{\mu\nu}m_j^{\nu\mu}~, \quad (m_im_jm_km_l) \equiv m_i^{\mu\nu}m_j^{\nu\rho}m_k^{\rho\sigma}m_l^{\sigma\mu} \,.}
where $\z_i$ and $p_i$ are the polarization vector and momenta of the $i$'th 10d graviton, respectively. In the conventions of \cite{DHoker:2005jhf} and \cite{Green:2008uj}, $\widehat{K} = 2^6 \cR^4$. See e.g. Appendix 9.A of \c{Green:1987mn} for the explicit form of $t_8$.

\ssec{Relation}

We now want to relate the preceding formulas to the flat space limit of the 4d $\cN=4$ superconformal Ward identity. Taking the large $s,t$ limit of \eqr{c2} (in which $u\rar -s-t$), we find
\e{c10}{ \lim_{s,t\rar\i}  \cM_p(s,t;\s,\tau) = {\Theta_4^{\rm flat}(s,t;\s,\tau)\o 16}\,\cMt_p(s,t)|_{s,t\rar\i}}
where
\e{}{\Theta_4^{\rm flat}(s,t;\s,\tau) \equiv (tu + ts\s+su\tau)^2}
\eqr{c10} actually holds for arbitrary $(p_1,p_2,p_3,p_4)$. Note that the right-hand side of \eqr{c10} is independent of $p$, unlike the analogous result in AdS$_7\times S^4$ \c{Chester:2018dga}. As shown in \c{Chester:2018dga},
\e{}{\widehat{K}\big|_{k_i\cdot y_i=0} = 4(y_1\cdot y_2)^2(y_3\cdot y_4)^2\Theta_4^{\rm flat}(s,t;\s,\tau)}

We now equate \eqr{c7} with \eqr{c10}. Given \eqr{a10}, this gives an $\cN=4$ SYM-based derivation of the overall kinematic factor in the type IIB string theory amplitude at arbitrary genus. Moreover, we read off the relation
\e{c13}{ \lim_{s,t\rar\i} \cMt_p(s,t) = 64 \,c(p) f(s,t) }
The constant $c(p)$ cancels out of a ratio of terms at different orders in the derivative expansion. 

At genus zero, $f(s,t)$ equals the Virasoro-Shapiro amplitude. This admits an analytic expansion in powers of $s,t$, 
\e{}{f^{(g=0)}(s,t) = 1+ \sum_{k=0}^\i \a'^{3+m+3n/2} f_{m,n}(s,t)}
The first term represents the 10d supergravity amplitude, while the rest contain the monomials $\s_2^m\s_3^n$, i.e. the $\p^{2k}\cR^4$ contributions with $2m+3n=2k$. Similarly, the $1/\l$ expansion of the reduced Mellin amplitude $\cMt_p$ at tree-level was written in \eqr{mpg0}. Using \eqr{c13}, we derive the relation between $f_{m,n}(s,t)$ and the flat space limit of $\cMt_{p|m,n}^{(g=0)}(s,t)$: 
\e{treereln}{{f_{m,n}(s,t) = {\l^{-(k+3)/2}\o 2^{k+5}(p+1)_{k+3}}\,{stu}\lim_{s,t\rar\i} {\cMt_{p|m,n}^{(g=0)}(s,t)}}}
where $2k=2m+3n$. This is the main formula of this Appendix. Note, in particular, the $p$-dependence. 

Taking $k=0$, corresponding to the $\cR^4$ term, on the IIB side we have, from the Virasoro-Shapiro amplitude \eqr{avs},
\e{}{f_{\cR^4}(s,t) = {\zeta_3\o 2^6} \a'^3 \s_3}
In the parameterization of \eqr{melling0}, $\cMt_{p|\cR^4}^{(g=0)}(s,t)=\a$. Plugging into \eqr{treereln} and using $\a'/L_{\rm AdS}^2 = 1/\sqrt{\l}$ yields
\e{}{\a = \zeta_3 (p+1)_3}
as reported in \eqr{alphabeta}. 

Taking $k=2$, corresponding to the $\p^4\cR^4$ term, and plugging
\e{}{f_{\p^4\cR^4}(s,t) = {\zeta_5\o 2^{10}}\a'^5\s_2\s_3}
and
\e{}{\cMt_{p|\p^{4}\cR^4}^{(g=0)}(s,t)={\b\s_2 + \b_1+\b_2 s}}
into \eqr{treereln} yields
\e{}{\b = {\zeta_5\o 8}(p+1)_5}
as reported in \eqr{alphabeta}. 

\section{Explicit form of solutions $S_0^{(q)}(z,\bar z)$}\label{appc}
In \eqr{lsums} we introduced the family of functions $S_L^{(q)}(z,\bar z)$ in terms of which the double-discontinuities under consideration can be written. In this appendix we give their explicit form for the simplest case $L=0$. For $q=0$ we obtain
\begin{equation}
S_0^{(0)}(z,\bar z) = \frac{P_0^{(5)}(z,\bar z) \log(1-\bar z) - P_0^{(5)}(\bar z,z) \log(1-z)+ P_1^{(5)}(z,\bar z)}{(z-\bar z)^7}
\end{equation}
where
\begin{eqnarray}
P_0^{(5)}(z,\bar z)&=&48 z^2 \bar{z}^2 (2-z-\bar{z}) \left(z^2+8 z \bar{z}-10 z+\bar{z}^2-10 \bar{z}+10\right)\\
P_1^{(5)}(z,\bar z)&=&-16 z^2 \bar{z}^2 (z-\bar{z}) \left(11 z^2+38 z \bar{z}-60 z+11 \bar{z}^2-60 \bar{z}+60\right)
\end{eqnarray}
From this expression we can generate all $S_0^{(q)}(z,\bar z)$ by a chain of differential operators
\begin{equation}
S_0^{(q+1)}(z,\bar z)={\cal D}_q S_0^{(q)}(z,\bar z) 
\end{equation}
where
\begin{eqnarray}
{\cal D}_q= \frac{(q+3) (z-1) z^2 \bar{z} (q \bar{z}-2 q-4)}{(z-\bar{z}) (q z \bar{z}-q z-q \bar{z}-2 z-2 \bar{z})} \partial_z -\frac{(q+3) z (\bar{z}-1) \bar{z}^2 (q z-2 q-4)}{(z-\bar{z}) (q z \bar{z}-q z-q \bar{z}-2 z-2 \bar{z})} \partial_{\bar z} 
\end{eqnarray}
This allows reconstruction of the contributions containing ${\cal R}^4$, as certain differential operator acting on $S_0^{(0)}(z,\bar z)$. Note that each action of $\mathcal{D}_q$ increases the power of $(z-\zb)^{-1}$ by two, so $S_0^{(q)}(z,\zb) \propto (z-\zb)^{-7-2q}$. A similar structure seems to be present for the functions $S_2^{(q)}(z,\bar z)$, but in this case the differential operator is much more involved.

\section{From the double-discontinuity to the anomalous dimension}\label{appd}     

In this appendix we derive the precise one-dimensional inversion integral that leads to the anomalous dimensions from the double-discontinuity for the present case. We follow closely \cite{Alday:2017vkk}. The idea is as follows. Consider the conformal block expansion of ${\cal H}(z,\bar z)$, keeping only the piece of the conformal block that can lead to a double-discontinuity at $\bar z=1$,
\begin{equation}
{\cal H}(z,\bar z) = \sum_{n,\ell} a_{n,\ell} (z \bar z)^{\frac{\tau_{n,\ell}}{2}} \frac{\bar z^{\ell+1} F_{\frac{\tau_{n,\ell}}{2}+1}(z)F_{\frac{\tau_{n,\ell}}{2}+\ell+2}(\bar z)}{\bar z-z} + \text{regular}
\end{equation}
where we have introduced the twist $\tau_{n,\ell} = \Delta_{n,\ell}-\ell$ and the regular contribution diverges at most as a single conformal block. From now on we will omit these contributions. In case of degenerate operators the sum over species is implicit, but this will not affect our analysis. The twist admits the following expansion
\begin{eqnarray}
\tau_{n,\ell}\!\!  \! \! \! \! \!  &=&\! \! \! \! \!\! \! \!  4+2n+ \frac{1}{c} \left( \gamma^{(g=0|{\rm sugra})}_{n,\ell}+\frac{1}{\lambda^{3/2}} \gamma^{(g=0|\cR^4)}_{n,\ell} + \cdots \right) + \frac{1}{c^2} \left( \gamma^{(g=1|{\rm sugra})}_{n,\ell}+\frac{1}{\lambda^{3/2}} \gamma^{(g=1|\cR^4)}_{n,\ell} + \cdots \right) \nonumber \\&+\cdots
\end{eqnarray}
The corrections $\gamma^{(g=1|\cR^4)}_{n,\ell}$ and beyond are our main concern in this paper. For concreteness we demonstrate with the $\l^{-3/2}$ term. Plugging this expansion into the above decomposition we find
\begin{equation}
\left. {\cal H}^{(g=1)}(z,\bar z) \right|_{\frac{1}{\lambda^{3/2}} \log z}= \frac{1}{2} \sum_{n,\ell} a^{(0)}_{n,\ell}  \gamma^{(g=1|\cR^4)}_{n,\ell} (z \bar z)^{2+n} \frac{\bar z^{\ell+1} F_{n+3}(z)F_{n+4+\ell}(\bar z)}{\bar z-z} 
\end{equation}
and similarly for other contributions. One may worry that at $\O(1/c^2)$ we may have other contributions to the  term proportional to $\log z$. However, this is not the case, since the solutions $\gamma^{(g=0,1)}_{n,\ell}$ are truncated in the spin, and hence cannot contribute to the double-discontinuity. Next we project on a given twist, using the projectors  
\begin{equation}
\frac{1}{2\pi i} \oint \frac{dz}{z} z^{n'} F_{n'+3}(z) \frac{F_{-2-n}(z)}{z^{n}} = \delta_{n,n'}
\end{equation}
which leads to
\begin{equation}
 \frac{1}{2} \sum_{\ell} a^{(0)}_{n,\ell}  \gamma^{(g=1|\cR^4)}_{n,\ell} \bar z^{2+n} \bar z^{\ell+1} F_{n+4+\ell}(\bar z)= \frac{1}{2\pi i} \oint \frac{dz}{z^{n+3}}F_{-2-n}(z) (\bar z-z)\left. {\cal H}^{(g=1)}(z,\bar z) \right|_{\frac{1}{\lambda^{3/2}} \log z}
\end{equation}
where the projection over a given $n$, not summed over, has been performed. The task is to invert $a^{(0)}_{n,\ell}  \gamma^{(g=1|\cR^4)}_{n,\ell}$ knowing the r.h.s. This problem has been solved in \cite{Alday:2016njk} in a $1/\ell$ expansion, while in \cite{Caron-Huot:2017vep} an elegant inversion formula was proposed. Both approaches are equivalent and it turns out that only the double-discontinuty of the r.h.s. is necessary. The final result can be repackaged as
\begin{eqnarray}\label{gammaddisc}
\frac{1}{2}a^{(0)}_{n,\ell}  \gamma^{(g=1|\cR^4)}_{n,\ell} &=& \frac{(7+2n+2\ell) r_{4+n+\ell}}{\pi^2  } \int_0^1 dt \int_0^1 d\bar z \frac{\bar z^{n+\ell+2}(t(1-t))^{n+\ell+3}}{(1-t \bar z)^{\ell+n+4}} \\
& & \times \frac{1}{2\pi i} \oint \frac{dz}{z^{n+3}}F_{-2-n}(z) {\rm dDisc} \left[(\bar z-z)\left. {\cal H}^{(g=1)}(z,\bar z) \right|_{\frac{1}{\lambda^{3/2}} \log z} \right] \nonumber
\end{eqnarray}
where $r_h \equiv \Gamma(h)^2/\Gamma(2h-1)$. The $t$ integral is just the integral representation of $F_{\ell+n+4}(\zb)$, but swapping the order of integration with $\zb$ leads to easier evaluation. This is the expression used in the body of the paper. It applies identically to higher vertices beyond $\cR^4$.

\bibliographystyle{utphys} 
\bibliography{stringy}

\providecommand{\href}[2]{#2}\begingroup\raggedright\begin{thebibliography}{10}

\bibitem{DHoker:1999kzh}
E.~D'Hoker, D.~Z. Freedman, S.~D. Mathur, A.~Matusis, and L.~Rastelli,
  ``{Graviton exchange and complete four point functions in the AdS / CFT
  correspondence},''
  \href{http://dx.doi.org/10.1016/S0550-3213(99)00525-8}{{\em Nucl. Phys.}
  {\bfseries B562} (1999) 353--394},
\href{http://arxiv.org/abs/hep-th/9903196}{{\ttfamily arXiv:hep-th/9903196
  [hep-th]}}.

\bibitem{Heemskerk:2009pn}
I.~Heemskerk, J.~Penedones, J.~Polchinski, and J.~Sully, ``{Holography from
  Conformal Field Theory},''
  \href{http://dx.doi.org/10.1088/1126-6708/2009/10/079}{{\em JHEP} {\bfseries
  10} (2009) 079},
\href{http://arxiv.org/abs/0907.0151}{{\ttfamily arXiv:0907.0151 [hep-th]}}.

\bibitem{Penedones:2010ue}
J.~Penedones, ``{Writing CFT correlation functions as AdS scattering
  amplitudes},'' \href{http://dx.doi.org/10.1007/JHEP03(2011)025}{{\em JHEP}
  {\bfseries 03} (2011) 025},
\href{http://arxiv.org/abs/1011.1485}{{\ttfamily arXiv:1011.1485 [hep-th]}}.

\bibitem{Alday:2014tsa}
L.~F. Alday, A.~Bissi, and T.~Lukowski, ``{Lessons from crossing symmetry at
  large N},'' \href{http://dx.doi.org/10.1007/JHEP06(2015)074}{{\em JHEP}
  {\bfseries 06} (2015) 074},
\href{http://arxiv.org/abs/1410.4717}{{\ttfamily arXiv:1410.4717 [hep-th]}}.

\bibitem{Aharony:2016dwx}
O.~Aharony, L.~F. Alday, A.~Bissi, and E.~Perlmutter, ``{Loops in AdS from
  Conformal Field Theory},''
  \href{http://dx.doi.org/10.1007/JHEP07(2017)036}{{\em JHEP} {\bfseries 07}
  (2017) 036},
\href{http://arxiv.org/abs/1612.03891}{{\ttfamily arXiv:1612.03891 [hep-th]}}.

\bibitem{Alday:2017xua}
L.~F. Alday and A.~Bissi, ``{Loop Corrections to Supergravity on $AdS_5 \times
  S^5$},'' \href{http://dx.doi.org/10.1103/PhysRevLett.119.171601}{{\em Phys.
  Rev. Lett.} {\bfseries 119} no.~17, (2017) 171601},
\href{http://arxiv.org/abs/1706.02388}{{\ttfamily arXiv:1706.02388 [hep-th]}}.

\bibitem{Alday:2016njk}
L.~F. Alday, ``{Large Spin Perturbation Theory for Conformal Field Theories},''
  \href{http://dx.doi.org/10.1103/PhysRevLett.119.111601}{{\em Phys. Rev.
  Lett.} {\bfseries 119} no.~11, (2017) 111601},
\href{http://arxiv.org/abs/1611.01500}{{\ttfamily arXiv:1611.01500 [hep-th]}}.

\bibitem{Caron-Huot:2017vep}
S.~Caron-Huot, ``{Analyticity in Spin in Conformal Theories},''
  \href{http://dx.doi.org/10.1007/JHEP09(2017)078}{{\em JHEP} {\bfseries 09}
  (2017) 078},
\href{http://arxiv.org/abs/1703.00278}{{\ttfamily arXiv:1703.00278 [hep-th]}}.

\bibitem{Alday:2017vkk}
L.~F. Alday and S.~Caron-Huot, ``{Gravitational S-matrix from CFT dispersion
  relations},''
\href{http://arxiv.org/abs/1711.02031}{{\ttfamily arXiv:1711.02031 [hep-th]}}.

\bibitem{Aprile:2017bgs}
F.~Aprile, J.~M. Drummond, P.~Heslop, and H.~Paul, ``{Quantum Gravity from
  Conformal Field Theory},''
  \href{http://dx.doi.org/10.1007/JHEP01(2018)035}{{\em JHEP} {\bfseries 01}
  (2018) 035},
\href{http://arxiv.org/abs/1706.02822}{{\ttfamily arXiv:1706.02822 [hep-th]}}.

\bibitem{Aprile:2017xsp}
F.~Aprile, J.~M. Drummond, P.~Heslop, and H.~Paul, ``{Unmixing Supergravity},''
  \href{http://dx.doi.org/10.1007/JHEP02(2018)133}{{\em JHEP} {\bfseries 02}
  (2018) 133},
\href{http://arxiv.org/abs/1706.08456}{{\ttfamily arXiv:1706.08456 [hep-th]}}.

\bibitem{Aprile:2018efk}
F.~Aprile, J.~Drummond, P.~Heslop, and H.~Paul, ``{The double-trace spectrum of
  $N=4$ SYM at strong coupling},''
\href{http://arxiv.org/abs/1802.06889}{{\ttfamily arXiv:1802.06889 [hep-th]}}.

\bibitem{Susskind:1998vk}
L.~Susskind, ``{Holography in the flat space limit},''
  \href{http://dx.doi.org/10.1063/1.1301570}{{\em AIP Conf. Proc.} {\bfseries
  493} no.~1, (1999) 98--112},
\href{http://arxiv.org/abs/hep-th/9901079}{{\ttfamily arXiv:hep-th/9901079
  [hep-th]}}.

\bibitem{Polchinski:1999ry}
J.~Polchinski, ``{S matrices from AdS space-time},''
\href{http://arxiv.org/abs/hep-th/9901076}{{\ttfamily arXiv:hep-th/9901076
  [hep-th]}}.

\bibitem{Okuda:2010ym}
T.~Okuda and J.~Penedones, ``{String scattering in flat space and a scaling
  limit of Yang-Mills correlators},''
  \href{http://dx.doi.org/10.1103/PhysRevD.83.086001}{{\em Phys. Rev.}
  {\bfseries D83} (2011) 086001},
\href{http://arxiv.org/abs/1002.2641}{{\ttfamily arXiv:1002.2641 [hep-th]}}.

\bibitem{Berkovits:2006vc}
N.~Berkovits, ``{New higher-derivative R**4 theorems},''
  \href{http://dx.doi.org/10.1103/PhysRevLett.98.211601}{{\em Phys. Rev. Lett.}
  {\bfseries 98} (2007) 211601},
\href{http://arxiv.org/abs/hep-th/0609006}{{\ttfamily arXiv:hep-th/0609006
  [hep-th]}}.

\bibitem{Green:1981yb}
M.~B. Green and J.~H. Schwarz, ``{Supersymmetrical String Theories},''
\href{http://dx.doi.org/10.1016/0370-2693(82)91110-8}{{\em Phys. Lett.}
  {\bfseries 109B} (1982) 444--448}.

\bibitem{Green:1997as}
M.~B. Green, M.~Gutperle, and P.~Vanhove, ``{One loop in eleven-dimensions},''
  \href{http://dx.doi.org/10.1016/S0370-2693(97)00931-3}{{\em Phys. Lett.}
  {\bfseries B409} (1997) 177--184},
  \href{http://arxiv.org/abs/hep-th/9706175}{{\ttfamily arXiv:hep-th/9706175
  [hep-th]}}.
[,164(1997)].

\bibitem{Russo:1997mk}
J.~G. Russo and A.~A. Tseytlin, ``{One loop four graviton amplitude in
  eleven-dimensional supergravity},''
  \href{http://dx.doi.org/10.1016/S0550-3213(97)00631-7,
  10.1016/S0550-3213(97)80012-0}{{\em Nucl. Phys.} {\bfseries B508} (1997)
  245--259},
\href{http://arxiv.org/abs/hep-th/9707134}{{\ttfamily arXiv:hep-th/9707134
  [hep-th]}}.

\bibitem{Green:1998by}
M.~B. Green and S.~Sethi, ``{Supersymmetry constraints on type IIB
  supergravity},'' \href{http://dx.doi.org/10.1103/PhysRevD.59.046006}{{\em
  Phys. Rev.} {\bfseries D59} (1999) 046006},
\href{http://arxiv.org/abs/hep-th/9808061}{{\ttfamily arXiv:hep-th/9808061
  [hep-th]}}.

\bibitem{Green:1999pu}
M.~B. Green, H.-h. Kwon, and P.~Vanhove, ``{Two loops in eleven-dimensions},''
  \href{http://dx.doi.org/10.1103/PhysRevD.61.104010}{{\em Phys. Rev.}
  {\bfseries D61} (2000) 104010},
\href{http://arxiv.org/abs/hep-th/9910055}{{\ttfamily arXiv:hep-th/9910055
  [hep-th]}}.

\bibitem{Green:1999pv}
M.~B. Green and P.~Vanhove, ``{The Low-energy expansion of the one loop type II
  superstring amplitude},''
  \href{http://dx.doi.org/10.1103/PhysRevD.61.104011}{{\em Phys. Rev.}
  {\bfseries D61} (2000) 104011},
\href{http://arxiv.org/abs/hep-th/9910056}{{\ttfamily arXiv:hep-th/9910056
  [hep-th]}}.

\bibitem{Green:2005ba}
M.~B. Green and P.~Vanhove, ``{Duality and higher derivative terms in M
  theory},'' \href{http://dx.doi.org/10.1088/1126-6708/2006/01/093}{{\em JHEP}
  {\bfseries 01} (2006) 093},
\href{http://arxiv.org/abs/hep-th/0510027}{{\ttfamily arXiv:hep-th/0510027
  [hep-th]}}.

\bibitem{Green:2006gt}
M.~B. Green, J.~G. Russo, and P.~Vanhove, ``{Non-renormalisation conditions in
  type II string theory and maximal supergravity},''
  \href{http://dx.doi.org/10.1088/1126-6708/2007/02/099}{{\em JHEP} {\bfseries
  02} (2007) 099},
\href{http://arxiv.org/abs/hep-th/0610299}{{\ttfamily arXiv:hep-th/0610299
  [hep-th]}}.

\bibitem{Green:2008uj}
M.~B. Green, J.~G. Russo, and P.~Vanhove, ``{Low energy expansion of the
  four-particle genus-one amplitude in type II superstring theory},''
  \href{http://dx.doi.org/10.1088/1126-6708/2008/02/020}{{\em JHEP} {\bfseries
  02} (2008) 020},
\href{http://arxiv.org/abs/0801.0322}{{\ttfamily arXiv:0801.0322 [hep-th]}}.

\bibitem{Goncalves:2014ffa}
V.~Goncalves, ``{Four point function of $\mathcal{N}=4$ stress-tensor multiplet
  at strong coupling},'' \href{http://dx.doi.org/10.1007/JHEP04(2015)150}{{\em
  JHEP} {\bfseries 04} (2015) 150},
\href{http://arxiv.org/abs/1411.1675}{{\ttfamily arXiv:1411.1675 [hep-th]}}.

\bibitem{Nirschl:2004pa}
M.~Nirschl and H.~Osborn, ``{Superconformal Ward identities and their
  solution},'' \href{http://dx.doi.org/10.1016/j.nuclphysb.2005.01.013}{{\em
  Nucl. Phys.} {\bfseries B711} (2005) 409--479},
\href{http://arxiv.org/abs/hep-th/0407060}{{\ttfamily arXiv:hep-th/0407060
  [hep-th]}}.

\bibitem{Dolan:2004mu}
F.~A. Dolan, L.~Gallot, and E.~Sokatchev, ``{On four-point functions of 1/2-BPS
  operators in general dimensions},''
  \href{http://dx.doi.org/10.1088/1126-6708/2004/09/056}{{\em JHEP} {\bfseries
  09} (2004) 056},
\href{http://arxiv.org/abs/hep-th/0405180}{{\ttfamily arXiv:hep-th/0405180
  [hep-th]}}.

\bibitem{Beem:2016wfs}
C.~Beem, L.~Rastelli, and B.~C. van Rees, ``{More ${\mathcal N}=4$
  superconformal bootstrap},''
  \href{http://dx.doi.org/10.1103/PhysRevD.96.046014}{{\em Phys. Rev.}
  {\bfseries D96} no.~4, (2017) 046014},
\href{http://arxiv.org/abs/1612.02363}{{\ttfamily arXiv:1612.02363 [hep-th]}}.

\bibitem{Arutyunov:2000py}
G.~Arutyunov and S.~Frolov, ``{Four point functions of lowest weight CPOs in
  N=4 SYM(4) in supergravity approximation},''
  \href{http://dx.doi.org/10.1103/PhysRevD.62.064016}{{\em Phys. Rev.}
  {\bfseries D62} (2000) 064016},
\href{http://arxiv.org/abs/hep-th/0002170}{{\ttfamily arXiv:hep-th/0002170
  [hep-th]}}.

\bibitem{Dolan:2001tt}
F.~A. Dolan and H.~Osborn, ``{Superconformal symmetry, correlation functions
  and the operator product expansion},''
  \href{http://dx.doi.org/10.1016/S0550-3213(02)00096-2}{{\em Nucl. Phys.}
  {\bfseries B629} (2002) 3--73},
\href{http://arxiv.org/abs/hep-th/0112251}{{\ttfamily arXiv:hep-th/0112251
  [hep-th]}}.

\bibitem{Rastelli:2016nze}
L.~Rastelli and X.~Zhou, ``{Mellin amplitudes for $AdS_5\times S^5$},''
  \href{http://dx.doi.org/10.1103/PhysRevLett.118.091602}{{\em Phys. Rev.
  Lett.} {\bfseries 118} no.~9, (2017) 091602},
\href{http://arxiv.org/abs/1608.06624}{{\ttfamily arXiv:1608.06624 [hep-th]}}.

\bibitem{Rastelli:2017udc}
L.~Rastelli and X.~Zhou, ``{How to Succeed at Holographic Correlators Without
  Really Trying},'' \href{http://dx.doi.org/10.1007/JHEP04(2018)014}{{\em JHEP}
  {\bfseries 04} (2018) 014},
\href{http://arxiv.org/abs/1710.05923}{{\ttfamily arXiv:1710.05923 [hep-th]}}.

\bibitem{Costa:2012cb}
M.~S. Costa, V.~Goncalves, and J.~Penedones, ``{Conformal Regge theory},''
  \href{http://dx.doi.org/10.1007/JHEP12(2012)091}{{\em JHEP} {\bfseries 12}
  (2012) 091},
\href{http://arxiv.org/abs/1209.4355}{{\ttfamily arXiv:1209.4355 [hep-th]}}.

\bibitem{Camanho:2014apa}
X.~O. Camanho, J.~D. Edelstein, J.~Maldacena, and A.~Zhiboedov, ``{Causality
  Constraints on Corrections to the Graviton Three-Point Coupling},''
  \href{http://dx.doi.org/10.1007/JHEP02(2016)020}{{\em JHEP} {\bfseries 02}
  (2016) 020},
\href{http://arxiv.org/abs/1407.5597}{{\ttfamily arXiv:1407.5597 [hep-th]}}.

\bibitem{Meltzer:2017rtf}
D.~Meltzer and E.~Perlmutter, ``{Beyond $a = c$: gravitational couplings to
  matter and the stress tensor OPE},''
  \href{http://dx.doi.org/10.1007/JHEP07(2018)157}{{\em JHEP} {\bfseries 07}
  (2018) 157},
\href{http://arxiv.org/abs/1712.04861}{{\ttfamily arXiv:1712.04861 [hep-th]}}.

\bibitem{Afkhami-Jeddi:2016ntf}
N.~Afkhami-Jeddi, T.~Hartman, S.~Kundu, and A.~Tajdini, ``{Einstein gravity
  3-point functions from conformal field theory},''
  \href{http://dx.doi.org/10.1007/JHEP12(2017)049}{{\em JHEP} {\bfseries 12}
  (2017) 049},
\href{http://arxiv.org/abs/1610.09378}{{\ttfamily arXiv:1610.09378 [hep-th]}}.

\bibitem{Kulaxizi:2017ixa}
M.~Kulaxizi, A.~Parnachev, and A.~Zhiboedov, ``{Bulk Phase Shift, CFT Regge
  Limit and Einstein Gravity},''
  \href{http://dx.doi.org/10.1007/JHEP06(2018)121}{{\em JHEP} {\bfseries 06}
  (2018) 121},
\href{http://arxiv.org/abs/1705.02934}{{\ttfamily arXiv:1705.02934 [hep-th]}}.

\bibitem{Costa:2017twz}
M.~S. Costa, T.~Hansen, and J.~Penedones, ``{Bounds for OPE coefficients on the
  Regge trajectory},'' \href{http://dx.doi.org/10.1007/JHEP10(2017)197}{{\em
  JHEP} {\bfseries 10} (2017) 197},
\href{http://arxiv.org/abs/1707.07689}{{\ttfamily arXiv:1707.07689 [hep-th]}}.

\bibitem{Gary:2009ae}
M.~Gary, S.~B. Giddings, and J.~Penedones, ``{Local bulk S-matrix elements and
  CFT singularities},''
  \href{http://dx.doi.org/10.1103/PhysRevD.80.085005}{{\em Phys. Rev.}
  {\bfseries D80} (2009) 085005},
\href{http://arxiv.org/abs/0903.4437}{{\ttfamily arXiv:0903.4437 [hep-th]}}.

\bibitem{Maldacena:2015iua}
J.~Maldacena, D.~Simmons-Duffin, and A.~Zhiboedov, ``{Looking for a bulk
  point},'' \href{http://dx.doi.org/10.1007/JHEP01(2017)013}{{\em JHEP}
  {\bfseries 01} (2017) 013},
\href{http://arxiv.org/abs/1509.03612}{{\ttfamily arXiv:1509.03612 [hep-th]}}.

\bibitem{Caron-Huot:2018kta}
S.~Caron-Huot and A.-K. Trinh, ``{All Tree-Level Correlators in
  AdS${}_5\times$S${}_5$ Supergravity: Hidden Ten-Dimensional Conformal
  Symmetry},''
\href{http://arxiv.org/abs/1809.09173}{{\ttfamily arXiv:1809.09173 [hep-th]}}.

\bibitem{Simmons-Duffin:2017nub}
D.~Simmons-Duffin, D.~Stanford, and E.~Witten, ``{A spacetime derivation of the
  Lorentzian OPE inversion formula},''
  \href{http://dx.doi.org/10.1007/JHEP07(2018)085}{{\em JHEP} {\bfseries 07}
  (2018) 085},
\href{http://arxiv.org/abs/1711.03816}{{\ttfamily arXiv:1711.03816 [hep-th]}}.

\bibitem{Liu:2018jhs}
J.~Liu, E.~Perlmutter, V.~Rosenhaus, and D.~Simmons-Duffin, ``{$d$-dimensional
  SYK, AdS Loops, and $6j$ Symbols},''
\href{http://arxiv.org/abs/1808.00612}{{\ttfamily arXiv:1808.00612 [hep-th]}}.

\bibitem{Virasoro:1969me}
M.~A. Virasoro, ``{Alternative constructions of crossing-symmetric amplitudes
  with regge behavior},''
\href{http://dx.doi.org/10.1103/PhysRev.177.2309}{{\em Phys. Rev.} {\bfseries
  177} (1969) 2309--2311}.

\bibitem{Alday:2016htq}
L.~F. Alday and A.~Bissi, ``{Unitarity and positivity constraints for CFT at
  large central charge},''
  \href{http://dx.doi.org/10.1007/JHEP07(2017)044}{{\em JHEP} {\bfseries 07}
  (2017) 044},
\href{http://arxiv.org/abs/1606.09593}{{\ttfamily arXiv:1606.09593 [hep-th]}}.

\bibitem{DHoker:2005vch}
E.~D'Hoker and D.~H. Phong, ``{Two-loop superstrings VI: Non-renormalization
  theorems and the 4-point function},''
  \href{http://dx.doi.org/10.1016/j.nuclphysb.2005.02.043}{{\em Nucl. Phys.}
  {\bfseries B715} (2005) 3--90},
\href{http://arxiv.org/abs/hep-th/0501197}{{\ttfamily arXiv:hep-th/0501197
  [hep-th]}}.

\bibitem{Gomez:2013sla}
H.~Gomez and C.~R. Mafra, ``{The closed-string 3-loop amplitude and
  S-duality},'' \href{http://dx.doi.org/10.1007/JHEP10(2013)217}{{\em JHEP}
  {\bfseries 10} (2013) 217},
\href{http://arxiv.org/abs/1308.6567}{{\ttfamily arXiv:1308.6567 [hep-th]}}.

\bibitem{DHoker:2015gmr}
E.~D'Hoker, M.~B. Green, and P.~Vanhove, ``{On the modular structure of the
  genus-one Type II superstring low energy expansion},''
  \href{http://dx.doi.org/10.1007/JHEP08(2015)041}{{\em JHEP} {\bfseries 08}
  (2015) 041},
\href{http://arxiv.org/abs/1502.06698}{{\ttfamily arXiv:1502.06698 [hep-th]}}.

\bibitem{DHoker:2017pvk}
E.~D'Hoker, M.~B. Green, and B.~Pioline, ``{Higher genus modular graph
  functions, string invariants, and their exact asymptotics},''
\href{http://arxiv.org/abs/1712.06135}{{\ttfamily arXiv:1712.06135 [hep-th]}}.

\bibitem{Chester:2018dga}
S.~M. Chester and E.~Perlmutter, ``{M-Theory Reconstruction from (2,0) CFT and
  the Chiral Algebra Conjecture},''
  \href{http://dx.doi.org/10.1007/JHEP08(2018)116}{{\em JHEP} {\bfseries 08}
  (2018) 116},
\href{http://arxiv.org/abs/1805.00892}{{\ttfamily arXiv:1805.00892 [hep-th]}}.

\bibitem{Chester:2018aca}
S.~M. Chester, S.~S. Pufu, and X.~Yin, ``{The M-Theory S-Matrix From ABJM:
  Beyond 11D Supergravity},''
  \href{http://dx.doi.org/10.1007/JHEP08(2018)115}{{\em JHEP} {\bfseries 08}
  (2018) 115},
\href{http://arxiv.org/abs/1804.00949}{{\ttfamily arXiv:1804.00949 [hep-th]}}.

\bibitem{DHoker:2005jhf}
E.~D'Hoker, M.~Gutperle, and D.~H. Phong, ``{Two-loop superstrings and
  S-duality},'' \href{http://dx.doi.org/10.1016/j.nuclphysb.2005.06.010}{{\em
  Nucl. Phys.} {\bfseries B722} (2005) 81--118},
\href{http://arxiv.org/abs/hep-th/0503180}{{\ttfamily arXiv:hep-th/0503180
  [hep-th]}}.

\bibitem{Green:1987mn}
M.~B. Green, J.~H. Schwarz, and E.~Witten, {\em {SUPERSTRING THEORY. VOL. 2:
  LOOP AMPLITUDES, ANOMALIES AND PHENOMENOLOGY}}.
\newblock 1988.
\newblock
\url{http://www.cambridge.org/us/academic/subjects/physics/theoretical-physics-and-mathematical-physics/superstring-theory-volume-2}.
\newblock

\end{thebibliography}\endgroup

\end{document}